\documentclass[reprint,NumberedRefs]{JASA}

\usepackage{algpseudocode}

\usepackage{siunitx}
\usepackage{booktabs}
\usepackage{enumitem}

\newcommand{\probEst}[1][i]{\ensuremath{P_{K = #1}}}
\newcommand{\modelT}{\ensuremath{T_i}}
\newcommand{\modelA}{\ensuremath{A_i}}
\newcommand{\modelN}{\ensuremath{N_0}}
\newcommand{\modelTprel}{\ensuremath{\widetilde{T}_i}}
\newcommand{\modelAprel}{\ensuremath{\widetilde{A}_i}}
\newcommand{\modelNprel}{\ensuremath{\widetilde{N}_0}}
\newcommand{\modelTest}{\ensuremath{\widehat{T}_i}}
\newcommand{\modelAest}{\ensuremath{\widehat{A}_i}}
\newcommand{\modelNest}{\ensuremath{\widehat{N}_0}}

\begin{document}

\title[Neural network for sound energy decay analysis]{Neural network for multi-exponential sound energy decay analysis}
\author{Georg Götz}
\email{georg.gotz@aalto.fi}
\author{Ricardo Falcón Pérez}
\author{Sebastian J. Schlecht}
\altaffiliation{Also at: Media Lab, Department of Art and Media, Aalto University, Otakaari 5, 00250 Espoo, Finland.}
\author{Ville Pulkki}
\affiliation{Aalto Acoustics Lab, Department of Signal Processing and Acoustics, Aalto University, P.O. Box 13100, 00076 Aalto, Finland}

\preprint{Götz et al.}	

\date{\today}

\begin{abstract}
An established model for sound energy decay functions (EDFs) is the superposition of multiple exponentials and a noise term. This work proposes a neural-network-based approach for estimating the model parameters from EDFs. The network is trained on synthetic EDFs and evaluated on two large datasets of over \num{20000} EDF measurements conducted in various acoustic environments. The evaluation shows that the proposed neural network architecture robustly estimates the model parameters from large datasets of measured EDFs\added{, while being lightweight and computationally efficient}. \deleted{It achieves an almost 30-fold speed-up compared to state-of-the-art methods.} An implementation of the proposed neural network is publicly available.  
\end{abstract}


\maketitle


\section{Introduction}
\label{sec:intro}
Room impulse responses (RIRs) are signals that describe the sound recorded by a receiver in a room in response to an impulsive excitation. They characterize how room, source, and receiver affect sound on the investigated transmission path, implicitly assuming a linear time-invariant system. The RIR decays gradually, thus causing the well-known acoustic phenomenon called reverberation \cite{Kuttruff2000RoomAcoustics}.

Due to pronounced fluctuations in the RIR, the energy-time curve may be hard to interpret and difficult to use in further analyses. Schroeder proposed a backwards integration procedure for obtaining smooth steady-state energy decay functions (EDFs) \cite{Schroeder1965BackwardsIntegration}, which are also called Schroeder decay \replaced{curves}{functions}. Schroeder decay \replaced{curves}{functions} are frequently used in architectural acoustics to calculate the reverberation time, which is the time until the sound energy in an enclosure has decreased by \SI{60}{\decibel}. The reverberation time is commonly determined for multiple frequency bands by fitting a straight line to the band-limited logarithmic EDF \cite{ISO3382RoomAcousticMeasurementPart1Performance,ISO3382RoomAcousticMeasurementPart2Ordinary}.

Measured RIRs are usually contaminated with noise, for instance, due to the measurement equipment, ambient sound, or quantization. The noise inevitably affects the EDF \cite{Chu1978ComparisonOfReverberationMeasurementsUsingSchroederAndAveraging} and causes errors when estimating the reverberation time based on a straight-line fit \cite{Morgan1997ParametricErrorAnalysisOfBackwardIntegrationLimit}. Several approaches were proposed to counter the effect of noise on the reverberation time estimation \cite{Chu1978ComparisonOfReverberationMeasurementsUsingSchroederAndAveraging,Lundeby1995UncertaintiesMeasurementsInRoomAcoustics,GuskiVorlaender2013MeasurementUncertaintiesRTcausedbyNoise,ISO3382RoomAcousticMeasurementPart1Performance}. Alternatively, Xiang \cite{Xiang1995EvaluationOfRTNonlinearRegression} and Karjalainen et al. \cite{Karjalainen2002EstimationOfModalDecayParameters} proposed to include an additional noise term in the model and perform a non-linear regression.

In coupled rooms or rooms with a considerably non-uniform absorption material distribution, sound energy can decay with multiple decay rates \cite{Eyring1931ReverberationTimeMeasurementsCoupledRooms,Hunt1939AnalysisOfSounDecayInRectRooms,Nilsson2004DecayInRoomsNonDiffuseSoundfieldCeilingTreatment,Kuttruff2000RoomAcoustics}. For this reason, Xiang and Goggans model EDFs with multiple exponential decays and a noise term \cite{Xiang2001EvaluationOfDecayTimesInCoupledSpacesBayesian}. 

This paper presents a neural-network-based approach for fitting multiple exponential decays and a noise term to an EDF. Despite being trained on a fully synthetic dataset, we show that such a neural network structure can robustly analyze real-world measurements. Fully synthetic training datasets can be easily generated and subsequently extended to different scenarios. Previous methods for multi-exponential sound energy decay analysis rely on iterative algorithms. Our approach has the advantage of being fully deterministic at \replaced{runtime}{inference time} and requiring no user-tuning, while being robust and computationally efficient. Therefore, the neural network structure is especially appealing for room acoustic analysis and modeling with machine-learning-based approaches, for which it is essential to achieve robust performance on large datasets without manual intervention. The proposed network is lightweight, allowing it to be implemented on mobile devices. Furthermore, the neural-network-based structure allows efficient up-scaling and parallelization on modern hardware with dedicated graphical processing units (GPUs).

Our contribution is threefold. Firstly, we present a \added{lightweight and computationally efficient} neural-network-based structure for sound energy decay analysis \replaced{that achieves an almost 30-fold speed-up compared to state-of-the-art methods, while maintaining a comparable fitting performance.}{that achieves a comparable fitting performance as state-of-the-art methods.} Secondly, we evaluate the proposed network and previous decay analysis approaches on two large datasets of more than \num{20000} EDFs. Thirdly, we provide an open-source decay analysis toolbox for MATLAB and Python, comprising the neural network structure and our implementations of the other evaluated multi-slope decay analysis methods.

The remainder of this paper is organized as follows. Section \ref{sec:problemformulation} states the problem formulation, and Section \ref{sec:priorwork} provides an overview of prior work. Section \ref{sec:proposedmethod} describes the proposed neural network and its training in detail. Section \ref{sec:Evaluation} presents an evaluation of the proposed network on two large datasets of real-world measurements and compares its performance to other state-of-the-art approaches. Section \ref{sec:Discussion} discusses the results. Section \ref{sec:Toolbox} details the publicly available decay analysis toolbox, and Section \ref{sec:conclusions} concludes the paper. 

\section{Problem formulation}
\label{sec:problemformulation}
Smooth energy decay functions (EDFs) can be obtained from RIRs via the backwards integration procedure proposed by Schroeder \cite{Schroeder1965BackwardsIntegration}. The EDF $d(t)$ of an RIR $h(t)$ is calculated by
\begin{linenomath*}
	\begin{equation}
		\label{eq:backwardsintegration}
		d(t) = \frac{1}{E} \ \sum_{\tau=t}^{L} h^2(\tau) \,, \quad \textrm{with} \quad E = \sum_{\tau=1}^{L} h^2(\tau) \,,
	\end{equation}
\end{linenomath*}
where $t$ is the sample index and $L$ is the number of samples in the EDF, which is also called the upper limit of integration (ULI).

EDFs can be modeled as a sum of multiple exponential decays and a noise term \cite{Xiang2001EvaluationOfDecayTimesInCoupledSpacesBayesian}. The model $d_K(t)$ of the EDF $d(t)$ is then given as \cite{Xiang2001EvaluationOfDecayTimesInCoupledSpacesBayesian,Xiang2005DecayInCoupledSpacesReliabilityAnalysisBayesianDecay} \explain{$-13.8$ instead of formulation with natural logarithm for readability}

\begin{linenomath*}
	\begin{equation}
		d_K(t) = N_0 (L - t) + \sum_{i=1}^{K} \modelA \left[ e^{\frac{-13.8 \cdot t} {f_\mathrm{s} \modelT}} - e^{\frac{-13.8 \cdot L} {f_\mathrm{s} \modelT}} \right] \,,
		\label{eq:MultiSlopeModel}
	\end{equation}
\end{linenomath*}
where \modelT{} and \modelA{} are the decay \replaced[time should be a better term, because a rate would be the entire fraction without the $t$]{rate}{time} and the amplitude of the $i$th exponential decay, respectively, \mbox{$\textrm{ln}(\cdot)$ denotes} the natural logarithm, \modelN{} is the amplitude of the noise term, \added[to make clear where the constant value in the exponents of Eq. (2) comes from]{the constant $-13.8 = \textrm{ln}(10^{-6})$ ensures that the sound energy has decayed to \SI{-60}{\decibel} after \modelT{} seconds,} $f_\mathrm{s}$ is the sampling frequency of the RIR, and $K$ is the model order, i.e., the number of exponential decays in the model. The constant second term in the square brackets accounts for the finite upper limit of integration and can be neglected for large~$L$~\cite{Xiang2005DecayInCoupledSpacesReliabilityAnalysisBayesianDecay}. 

Estimating the parameters \modelT{} and \modelA{} is a crucial task for various problems in room acoustics. For example, the reverberation time can be determined by estimating the parameter $T_1$ for an EDF model with $K = 1$. Decay models of higher order have successfully been used to measure the absorption coefficients of materials \cite{BalintMuralterEtAl2019BayesianDecayTimeEstimationAbsorptionMeasurements} or characterize the sound decay of coupled spaces \cite{Pu2011SoundDecayPatternsEnergyFeedbackCoupled,SuGul2019DiffusionEquationModelingSoundEnergyFlowCoupled}.

\section{Prior work}
\label{sec:priorwork}
\subsection{Sound energy decay analysis}
Previous approaches for estimating the parameters \modelT{}, \modelA{}, and \modelN{} differ mainly regarding the underlying model order. For model order $K = 1$, linear regression is commonly used to determine the reverberation time as a straight line fit to the band-limited logarithmic EDF \cite{ISO3382RoomAcousticMeasurementPart1Performance,ISO3382RoomAcousticMeasurementPart2Ordinary}. In this case, the noise term of the model is neglected, i.e., $N_0 = 0$. To get accurate estimates for $T_1$, the effect of the noise has to be countered by noise subtraction \cite{Chu1978ComparisonOfReverberationMeasurementsUsingSchroederAndAveraging} or truncation of the RIR before backwards integration \cite{Lundeby1995UncertaintiesMeasurementsInRoomAcoustics,GuskiVorlaender2013MeasurementUncertaintiesRTcausedbyNoise,ISO3382RoomAcousticMeasurementPart1Performance}. The noise term \modelN{} can be included in the model by using non-linear regression \cite{Xiang1995EvaluationOfRTNonlinearRegression, Karjalainen2002EstimationOfModalDecayParameters}.

The sound decay of coupled rooms or rooms with considerably non-uniform absorption material distribution can usually not be modeled with a single decay rate \cite{Eyring1931ReverberationTimeMeasurementsCoupledRooms,Hunt1939AnalysisOfSounDecayInRectRooms,Nilsson2004DecayInRoomsNonDiffuseSoundfieldCeilingTreatment,Kuttruff2000RoomAcoustics}. In such cases, model orders $K > 1$ need to be considered. Xiang and Goggans proposed a Bayesian framework to determine the model parameters \modelT{}, \modelA{}, and \modelN{} for $K \geq 1$ \cite{Xiang2001EvaluationOfDecayTimesInCoupledSpacesBayesian}. The Bayesian formulation can also determine the most probable model order $K$ given the measured EDF \cite{XiangEtAl2011BayesianCharacterizationOfMultiSlopeDecays}. Numerous works have advanced the approach by investigating more accurate and efficient algorithms for estimating the parameters or determining the model order \cite{JasaXiang2009EfficientEstimationDecayParamsSliceSampling,XiangEtAl2011BayesianCharacterizationOfMultiSlopeDecays, BalintMuralterEtAl2019BayesianDecayTimeEstimationAbsorptionMeasurements}.

Previous studies have evaluated the performance of single-slope EDF fitting software \cite{Cabrera2016CalculatingReverberationTimeFromIRSoftwareComparison,AlvarezMorales2016AcousticCharacterisationDifferentRoomAcousticSoftware,Katz2004RoundRobinRoomAcousticIRAnalysisSoftware}. Katz \cite{Katz2004RoundRobinRoomAcousticIRAnalysisSoftware} used a single measured lecture theater RIR, which was also used in a later study by Cabrera et al. \cite{Cabrera2016CalculatingReverberationTimeFromIRSoftwareComparison} together with artificial single-slope responses. Álvarez-Morales et al. \cite{AlvarezMorales2016AcousticCharacterisationDifferentRoomAcousticSoftware} studied the software behavior on a slightly larger set of RIRs ($15$ receiver positions $\times$ $2$ source positions, measured in a single auditorium). While Katz \cite{Katz2004RoundRobinRoomAcousticIRAnalysisSoftware} still reported considerable differences among reverberation time estimation software in 2004, the later studies \cite{Cabrera2016CalculatingReverberationTimeFromIRSoftwareComparison,AlvarezMorales2016AcousticCharacterisationDifferentRoomAcousticSoftware} found that reverberation time is consistently estimated within perceptual limits by the more recent software implementations in most cases. In all studies, the most inconsistent reverberation time estimates were obtained for low-frequency bands \cite{Cabrera2016CalculatingReverberationTimeFromIRSoftwareComparison,AlvarezMorales2016AcousticCharacterisationDifferentRoomAcousticSoftware,Katz2004RoundRobinRoomAcousticIRAnalysisSoftware}. 


\subsection{Convolutional neural networks in acoustics}
\begin{figure*}[ht]
	\begin{center}\includegraphics[width=\textwidth]{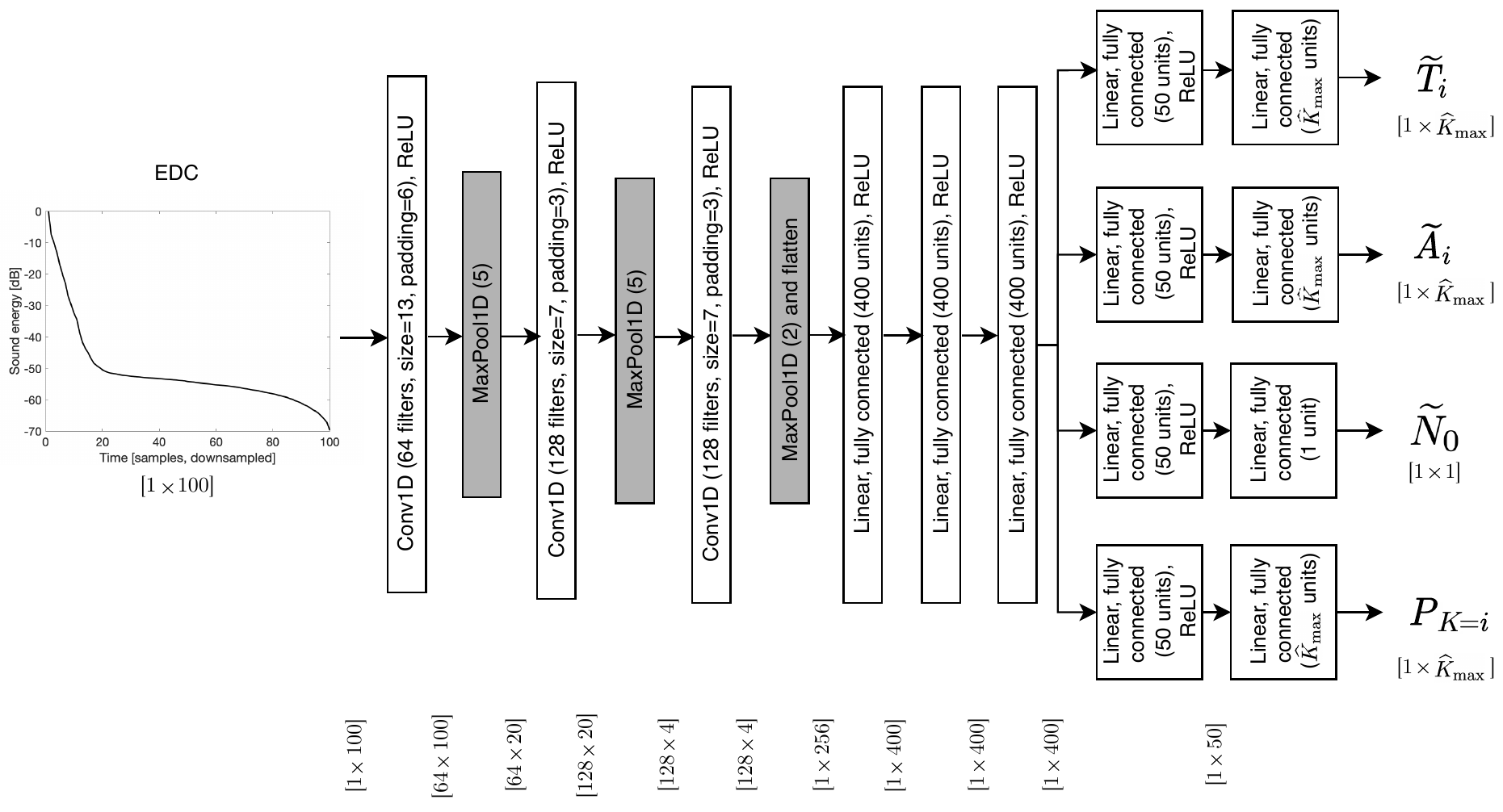}
		\caption{\label{fig:decayfitnet}An example of the proposed neural network structure for the parameter estimation of the sound decay model in Eq. (\ref{eq:MultiSlopeModel}). Throughout this paper, we refer to this particular architecture as DecayFitNet. The network outputs are preliminary values for the decay times \modelTprel{}, decay amplitudes \modelAprel{}, and noise term \modelNprel. Additionally, the network returns the values \probEst{}, which quantify the EDF model order prediction. The input length and maximum model order for the current implementation are $M=100$ samples and $\widehat{K}_\textrm{max} = 3$, respectively.}
	\end{center}
\end{figure*}
The recent advances in machine learning and artificial intelligence have pushed the state of the art in many different fields. Machine learning has traditionally been of interest in speech recognition \cite{ALAM2020302}, natural language processing \cite{young2017recent}, and music information retrieval \cite{Purwins2019}. More recent applications in acoustics include source localization and tracking \cite{grumiaux_2021:SurveySoundSource}, blind room acoustic parameter estimation \cite{ACEchallenge}, sound field scattering from 3D objects \cite{fan2020fast,Wirler2021Scattering}, and the inverse problem of object geometry regression from the scattered sound field \cite{9414743}. For more details, the reader is referred to the thorough review by Bianco et al. \cite{bianco_2019:MachineLearningAcoustics}. Although the applications, models, and task definitions vary considerably, most of the previously mentioned works share a common approach: a machine learning model, typically a neural network, extracts features and predicts the parameters of a parametric model. 

Most of these applications can be divided into either classification or regression tasks, where the main difference is the output domain \cite{GoodBengCour16}. For classification, the goal is to select one or several categories, given a specific input. For regression, the goal is to predict one or several continuous values. The study by Fernández-Delgado et al. \cite{FERNANDEZDELGADO201911} includes a detailed analysis of many regression methods and applications. Depending on the network architecture and training procedure, some tasks can be formulated as regression or classification tasks. In this paper, the decay parameter prediction is treated as a regression task, where we estimate the decay parameters as continuous values. On the other hand, the model order prediction is treated as a classification task with the categories ``1 active decay slope'', ``2 active decay slopes'', and so forth.


\section{Proposed method}
\label{sec:proposedmethod}
In this paper, we propose a neural network structure for estimating the parameters \modelT{}, \modelA{}, and \modelN{} of the model \mbox{in Eq. (\ref{eq:MultiSlopeModel}).} We show that such a network can be trained on a fully-synthetic training dataset to make predictions on real-world measurements. In the following section, we will first describe an example of such a network architecture. Subsequently, we provide details on the synthetic training dataset, the utilized loss functions, and the hyperparameters that can be used to train the network. 

\subsection{Network architecture}
\label{subsec:networkarchitecture}
\explain{Erroneous ReLU activations on final layer have been deleted from Fig. 1} An example of the proposed neural network structure for estimating the sound decay parameters of the model in \mbox{Eq. (\ref{eq:MultiSlopeModel})} is depicted in Figure \ref{fig:decayfitnet}. We refer to this particular architecture as DecayFitNet in the remainder of this paper. The network takes EDFs as its input and returns preliminary estimates \modelTprel{}, \modelAprel{}, and \modelNprel{}  for the decay \replaced[time should be a better term, because a rate would be the entire fraction in Eq. (2) without the $t$]{rates}{times}, decay amplitudes, and noise term, respectively. Furthermore, it outputs the values \probEst{}, which quantify the model order prediction. For example, by applying the logistic function
\begin{equation}
f(x) = \frac{1}{1 + e^{-x}} \,,
\end{equation}
we could get the probabilities $f(\probEst{}) = [0.75, 0.15, 0.10]$, which are bound between~$0$~and~$1$, and indicate that the network predicts the model order $\widehat{K}$ to be $ 1, 2$, or $3$ with the probabilities $0.75$, $0.15$, and $0.10$, respectively. In the current implementation, we restrict the maximum model order to $\widehat{K}_\textrm{max}=3$, although higher model orders could also be supported in the future. 


\deleted{Erroneous ReLU activations of final layers have been deleted from Fig. 1.} 

The network consists of a common base and individual output branches for the different estimated parameters. The base contains a sequence of three one-dimensional convolutional layers with intermediate max-pooling layers \cite{Scherer2010MaxPool} along the time-axis and three fully-connected layers. The output branches consist of two fully-connected layers each. Rectified Linear Units (ReLU) \cite{Nair2010ReLUImproveRestrictedBoltzmannMachines} are used as activation functions after each network layer, excluding the output layers.

This type of network architecture is ubiquitous for many different applications and tasks. In theory, a network that only consists of fully-connected layers and nonlinearities can be seen as a universal approximator \cite{Pinkus1999ApproximationTO}. Such networks are also called multi-layer perceptron (MLP). In practice, the flexibility of MLPs is limited by the amount of training data, size of the network, and training procedure. Most recent deep learning approaches favor convolutional layers to reduce the number of weights needed \cite{GoodBengCour16}. Furthermore, models with subsequent blocks of convolutional layers, nonlinearities, and pooling layers are biologically inspired by the visual processing system found in many living beings \cite{lecun95convolutional}. The convolutional layers act as feature extractors by learning filters that process the input signal in such a way as to maximize the information required for the task. The max-pooling layers behave like a downsampling operation that focuses on the most active features, thus reducing dimensionality and redundancies. Consecutive blocks represent progressively higher-level features of the input signal. The shared fully-connected layers recombine the extracted features in a non-linear way. Finally, the task-specific fully-connected layers act as independent regressors.

Using a shared common core for all tasks has several advantages over using independent networks for each task. First, this reduces the system complexity for both training and inference, as the number of computations is reduced. More importantly, the shared core will tend to learn useful features across all tasks and not overfit any of them \cite{bilen2016}. This property can improve generalization and robustness to noisy data.  

\subsection{Pre-processing}
\label{subsec:preprocessing}
A neural-network-based structure for decay analysis requires some pre-processing steps. In the following, we detail the pre-processing steps for our example implementation \mbox{DecayFitNet.} 

As the network requires a fixed-length input, we propose three pre-processing steps on the EDF $d(t)$ to be analyzed:
\begin{enumerate}[label=(\roman*)]
	\item The last \SI{5}{\percent} of the EDF are excluded, i.e., the samples $d(t > 0.95 L)$ are discarded. This step is motivated by the statistical uncertainty of the last EDF samples. This uncertainty is inherent to the Schroeder backwards integration method because, at the end of the EDF, only few RIR samples are integrated.
	\item The result is resampled to a fixed length $M$. In our example implementation \mbox{DecayFitNet}, we set $M=100$ samples. We apply fractional resampling.
	\item The EDF is converted into logarithmic scale (in \si{\decibel}) and normalized with the biggest absolute EDF sample value of the entire training dataset. The normalization ensures that every EDF sample of the training dataset lies in the interval $\left[-1,\, 1\right]$. The required normalization factor must be saved to a normalization file after the training procedure is completed. This way, the normalization can also be applied during inference.
\end{enumerate}

\subsection{Post-processing of estimates}
The preliminary parameter estimates are processed with the following output transformations to yield the final parameter estimates \modelTest{},  \modelAest{}, and \modelNest{}:
\begin{linenomath*}
	\begin{subequations}
	\begin{align}
		\widehat{K} =&\ \textrm{argmax}_i(\probEst{}) \\
		\modelTest =&\ \frac{\modelTprel^2 + 1}{M} \ \frac{L}{f_\textrm{s}} \\
		\modelAest =&\ 
		\begin{cases} 
			\modelAprel^2  &\textrm{ for}\ i \leq \widehat{K}\\
			0  &\textrm{ for}\ i > \widehat{K}
		\end{cases} \\
		\modelNest =&\ \frac{M}{L} \ 10^{-\modelNprel} \,.
	\end{align}
	\end{subequations}
\end{linenomath*}

We use these transformations to ensure that the final predictions $\modelTest > \SI{0}{\second}$, $\modelAest \geq 0$, and to achieve better numerical stability during the training while covering a large dynamic range of background noise levels \modelNest{}. While \modelTest{} is the estimated decay time in seconds, the preliminary estimate \modelTprel{} is the decay time in samples. Therefore, the value $\frac{\modelTprel^2 + 1}{M}$ is the decay time relative to the \added{neural network} input length~$M$, where squaring and adding~$1$~results in $\modelTest{} > 0$, thus avoiding division by zero in the exponential terms \mbox{[c.f. Eq. (\ref{eq:MultiSlopeModel})].} Due to the resampling in the pre-processing stage, the decay \replaced[time should be a better term, because a rate would be the entire fraction in Eq. (2) without the $t$]{rate}{time} estimates must be readjusted to the original time scale by multiplying with~$\frac{L}{f_\textrm{s}}$. For the same reason, the noise value predictions are scaled by~$\frac{M}{L}$. The amplitudes of all exponential terms that have a higher order than the predicted model order $\widehat{K}$ are set to zero, thus effectively removing their contribution to the predicted EDF fit. Although the amplitude values \modelA{} can cover a large dynamic range, our preliminary experiments found that the training converges to better results when we predict \modelAest{} on a linear scale, as opposed to the logarithmic scale that we use to predict the noise~values~\modelNest. 

The estimated fit $\widehat{d}_{\widehat{K}}(t)$ is obtained by inserting the network predictions \modelTest, \modelAest, and \modelNest{} into Eq. (\ref{eq:MultiSlopeModel}).

\subsection{Synthetic training dataset}
\label{subsec:dataset}
A large quantity of EDFs with various combinations of the decay parameters \modelT, \modelA, and~\modelN{}~[c.f.~Eq.~(\ref{eq:MultiSlopeModel})] is required to train the proposed neural network structure. Synthetic EDFs are an efficient way of collecting many different EDFs that cover the variety of real-world measurements. In the following, we want to detail how such a training dataset can be synthesized.

A large fully-synthetic dataset of \num{150000} EDFs was generated to train the DecayFitNet. It can be split into three equally-sized subsets, consisting of EDFs with model orders of $1$, $2$, and $3$, respectively. The data generation included three steps. Firstly, we randomly assigned values for the decay parameters \modelT, \modelA, and \modelN. Secondly, for model orders bigger than $1$, we checked if the drawn values were different enough to produce a proper multi-slope EDF and redraw values if necessary. Lastly, Gaussian noise was octave-band filtered, shaped, and backwards-integrated to obtain a synthetic EDF with the previously drawn \modelT, \modelA, and \modelN{} values. Details about these steps are elaborated in the following.

Values for the decay \replaced[time should be a better term, because a rate would be the entire fraction in Eq. (2) without the $t$]{rates}{times} \modelT{} were drawn from the uniform distribution 
\begin{linenomath*}
	\begin{equation}
		\modelT \in \mathcal{U}(0.1\ T_\textrm{EDF},\, 1.5\ T_\textrm{EDF})\,,
	\end{equation} 
\end{linenomath*}
where $T_\textrm{EDF} = \frac{L}{f_\textrm{s}}$. In other words, the drawn \modelT{} values are between \SIrange{10}{150}{\percent} of the input EDF length $T_\textrm{EDF}$. For our training dataset, we set $T_\textrm{EDF} = \SI{10}{\second}$ and $f_\textrm{s} = \SI{48}{\kilo\Hz}$. Due to the resampling of arbitrary length EDFs during inference, our dataset allows predicting different \modelT{} value ranges for different input EDF lengths. For instance, for a \SI{2.5}{\second} long input EDF, the network can predict \modelT{} values with lower and upper limits corresponding to $0.1 \times \SI{2.5}{\second} = \SI{0.25}{\second}$ and $1.5 \times \SI{2.5}{\second} = \SI{3.75}{\second}$, respectively. In other words, the resampling allows our network to operate on various reverberation time ranges, depending on the length of the input EDF. After drawing the values, they were ordered, such that $T_{i+1} \geq T_i$.

In our training dataset we wanted to cover an extensive dynamic range. Therefore, we assigned
\begin{linenomath*}
	\begin{equation} \modelN = 10^{a_\textrm{noise}}, \quad \textrm{with} \quad a_\textrm{noise} \in \mathcal{U}(-14, -3)\,,
	\end{equation}
\end{linenomath*}
corresponding to noise values between \SI{-140}{\decibel} and \SI{-30}{\decibel}.

As the decay amplitudes \modelA{} typically cover a large dynamic range as well, we assigned the values as 
\begin{linenomath*}
	\begin{equation}
		\modelA{} = 10^{a},\quad \textrm{with}\quad  a \in \mathcal{U}(-3,\, 0)\,, 
	\end{equation} 
\end{linenomath*}
corresponding to amplitude values between \SI{-30}{\decibel} and \SI{0}{\decibel}. The \modelA{} values were normalized, such that they sum up to unity, i.e., $\sum\limits_{i=1}^{K} \modelA = 1$. Finally, they were ordered, such that~$A_{i} \geq A_{i+1}$.

If the desired model order $K$ was bigger than $1$, the initial \modelT{} and \modelA{} values were checked for a sufficient multi-slope characteristic. This step was crucial because the random assignment could result in almost identical \modelT{} and \modelA{} values for the different slopes, thus generating a single-slope EDF, although a multi-slope EDF was desired. Therefore, we introduced the constraints 
\begin{linenomath*}
	\begin{subequations}
	\begin{align}
		T_{i+1} &\geq 1.5\ T_i \label{eq:tconstraint}\\
		\frac{A_{i}}{A_{i+1}} &\geq 10^{\frac{3}{K}} \,. \label{eq:aconstraint}
	\end{align}
	\end{subequations}
\end{linenomath*}
The constraint of Eq. (\ref{eq:tconstraint}) ensured that the different slopes have considerably different decay rates. By applying the constraint of Eq. (\ref{eq:aconstraint}), we aimed to distribute the amplitude values \modelA{} over the entire range from \SIrange{-30}{0}{\decibel}, thus preventing very similar values. New values for \modelT{} and \modelA{} were randomly drawn from the above distributions until both constraints were fulfilled. A visual inspection of some resulting EDFs showed that both constraints together ensured a distinct multi-slope character. 

Instead of directly inserting the resulting \modelT{}, \modelA{}, and \modelN{} values into Eq. (\ref{eq:MultiSlopeModel}), the final synthetic EDF was generated by applying the backwards integration [c.f. Eq. (\ref{eq:backwardsintegration})] on decaying Gaussian noise. This additional step introduced small random fluctuations into the synthetic EDFs to improve generalization after training the neural network. Four steps were necessary to obtain the final synthetic EDF. Firstly, the \modelA{} and \modelN{} values had to be scaled to account for the backwards integration
\begin{linenomath*}
	\begin{subequations}
	\begin{align}
		A_{i, \textrm{synth}} &= 13.8 \cdot T_\textrm{EDF} \ \frac{A_i }{T_i} \\
		N_{0, \textrm{synth}} &= M \ \modelN \,.
	\end{align}
	\end{subequations}
\end{linenomath*}
Secondly, a synthetic energy response was generated as
\begin{linenomath*}
	\begin{equation}
		\begin{aligned}
			h_\textrm{synth}^2(t) =\ & N_{0, \textrm{synth}} \cdot (f_\textrm{s}\cdot T_\textrm{EDF} - t)\cdot g_0^2(t) \\ &+ \sum_{i=1}^{K} A_{i, \textrm{synth}}\cdot g_i^2(t)\cdot e^{\frac{-13.8 \cdot t} {f_\mathrm{s} \modelT}} \,,
		\end{aligned}	
\end{equation}
\end{linenomath*}
where $g_0(t), g_1(t), ..., g_K(t)$ is Gaussian noise that is filtered with a random octave-band filter, and re-normalized to zero mean and unit variance. For the data generation we assumed $f_\textrm{s} = \SI{48}{\kilo\Hz}$ and $T_\textrm{EDF} = \SI{10}{\second}$. Thirdly, the response $h_\textrm{synth}(t)$ was backwards-integrated according to \mbox{Eq. (\ref{eq:backwardsintegration}).} Lastly, just as described in Section \ref{subsec:preprocessing}, the last \SI{5}{\percent} of the EDF samples were discarded, and the result was resampled to a length of $M=100$ samples. 

\subsection{Loss function}
We propose to use a loss function consisting of three parts for training the proposed neural network structure.\\
The first part is the \textit{EDF loss}, defined as the mean absolute error (MAE) between the analyzed EDF $d_\textrm{dB}(t)$ and the estimated fit $\widehat{d}_{\widehat{K}, \textrm{dB}}(t)$
\begin{linenomath*}
	\begin{equation}
		\mathcal{L}_\textrm{EDF} = \frac{1}{M} \sum_{t=0}^{M-1} \lvert d_\textrm{dB}(t) - \widehat{d}_{\widehat{K}, \textrm{dB}}(t) \rvert \,,
	\end{equation}
\end{linenomath*}
where the last \SI{5}{\percent} of EDC samples are excluded for both EDFs (c.f. Section \ref{subsec:preprocessing}), both EDFs are converted to a logarithmic scale (in \si{\decibel}), and $\lvert . \rvert$ denotes the absolute value.\\
The second part of the loss is the \textit{noise loss}, defined as the absolute error between the ground truth and the estimated noise exponent
\begin{linenomath*}
	\begin{equation}
		\mathcal{L}_\textrm{noise} = \lvert \textrm{log}_{10}(N_{0}) - \textrm{log}_{10}(\widehat{N}_{0})\rvert \,.
	\end{equation}
\end{linenomath*}
The third part of the loss is the \textit{model order loss}, defined as the cross-entropy loss
\begin{linenomath*}
	\begin{equation}
		\mathcal{L}_\textrm{order} = -\probEst[\widehat{K}] + \textrm{ln} \left(\sum_{i=1}^{\widehat{K}_\textrm{max}} e^{\probEst} \right) \,.
		\label{eq:crossentropy}
	\end{equation}
\end{linenomath*}
In Eq. (\ref{eq:crossentropy}), \probEst{} quantifies which probability the network assigns to the model order $i$, where $K$ is the true model order. We include the model order loss to teach the network to predict the correct number of slopes in an EDF. This measure should prevent the network from outputting multiple similar slopes in cases where a single slope would fit the EDF sufficiently well.\\
Finally, the total loss $\mathcal{L}$ for training the proposed neural network is
\begin{linenomath*}
	\begin{equation}
		\mathcal{L} = \mathcal{L}_\textrm{EDF} + \mathcal{L}_\textrm{noise} + \mathcal{L}_\textrm{order} \,.
	\end{equation}
\end{linenomath*}

\subsection{Training}
We train the DecayFitNet for $200$ epochs using the Adam optimizer \cite{KingmaBa2015Adam} with an initial learning rate of $1 \times 10^{-3}$ and a weight decay \cite{Loshchilov2019DecoupledWeightDecay} of $3 \times 10^{-4}$. Additionally, we apply cosine annealing with warm restarts \cite{Loshchilov2017CosAnnealingWarmRestarts} using a schedule of $40$ epochs between the restarts.

\section{Evaluation}
\label{sec:Evaluation}
In our evaluation, we use the DecayFitNet on two large datasets and compare its EDF fitting performance with a publicly available toolbox and our own implementation of the Bayesian decay analysis framework. The following section describes the details of our evaluation.

\subsection{Baseline methods}
We compare the DecayFitNet with two other decay analysis approaches. The first baseline method is based on a non-linear regression model of a single exponential and a noise term \cite{Karjalainen2002EstimationOfModalDecayParameters}. The second baseline method is based on slice sampling for Bayesian decay analysis \cite{JasaXiang2009EfficientEstimationDecayParamsSliceSampling}. Details about the implementation of both methods are presented in the following.

\subsubsection{Non-linear regression} 
This baseline method uses the publicly available toolbox implemented by Karjalainen et al. \cite{Karjalainen2002EstimationOfModalDecayParameters}. In initial experiments, we found that the performance of the toolbox depends considerably on the choice of the fitting scale. This issue was already observed by Karjalainen et al., which is why they proposed to fit the EDF $d(t)$ on a power scale \cite{Karjalainen2002EstimationOfModalDecayParameters}. This means that the non-linear regression is carried out on the scaled EDF $d_\textrm{scale}(t) = d^s(t)$, thus adding the adjustable hyperparameter $s$ to the method. In our evaluation, we use $s = 0.5$ as suggested by the developers of the toolbox \cite{Karjalainen2002EstimationOfModalDecayParameters}. Additionally, we use an improved variant, where a grid search is carried out over the interval $s \in \left[ 0.2,\ 0.8 \right]$ to find the best fit regarding the mean squared \mbox{error (MSE).} In both variants, we only use the EDF below \SI{-5}{\decibel}, which is common practice for reverberation time estimation.

\subsubsection{Bayesian decay analysis}
For the second baseline method of our evaluation, we implemented a slice-sampling-based Bayesian decay analysis \cite{JasaXiang2009EfficientEstimationDecayParamsSliceSampling} in MATLAB and Python. The code is contained in our Decay Analysis Toolbox~(c.f.~Section~\ref{sec:Toolbox}). Our implementation is based on the fully parameterized Bayesian formulation using the likelihood $\ell$ defined as \cite{JasaXiang2009EfficientEstimationDecayParamsSliceSampling}
\begin{linenomath*}
	\begin{equation}
\ell = \Gamma\left(\frac{L}{2}\right) \frac{(2\pi \mathcal{E})^{-L/2}}{2} \,,
	\end{equation}
\end{linenomath*}
where $\Gamma(\cdot)$ is the gamma function, $L$ is the number of EDF samples, and $\mathcal{E}$ quantifies the error between the measured EDF $d(t)$ and the model $d_K(t)$ as defined in Eq. \eqref{eq:MultiSlopeModel}:
\begin{linenomath*}
	\begin{equation}
\mathcal{E} = \frac{1}{2} \sum_{t=1}^{L} \left[ d(t) - d_K(t)\right]^2 \,.
	\end{equation}
\end{linenomath*}
No prior information about the parameter values \modelT, \modelA, and \modelN{} is available before the decay analysis. Therefore, we assign a uniform prior and estimate the model parameters by maximizing the likelihood $\ell$ over the parameter space. 

For this purpose, we apply the slice sampling algorithm \cite{JasaXiang2009EfficientEstimationDecayParamsSliceSampling}, because a grid search over all parameter combinations would be computationally infeasible. In our analysis, we let the slice sampling algorithm run for $50$ iterations and restrict the search space as follows
\begin{linenomath*}
	\begin{subequations}
	\begin{alignat}{4}
		&T_i &&\in\, \mathcal{U}(\SI{0.1}{\second},\, \SI{3.5}{\second})  &&\\
		&A_i &&= 10^a,&&\textrm{with}\quad  a  &&\in\,   \mathcal{U}(-3,\, 0) \\
		&N_0 &&= 10^{a_\textrm{noise}},&&\textrm{with}\quad  a_\textrm{noise}  &&\in\,   \mathcal{U}(-10,\, -2) \,,
	\end{alignat}
	\end{subequations}
\end{linenomath*}
where each dimension is discretized into $100$ points (equally-spaced in $T_i$, logarithmically-spaced in $A_i$, and $N_0$). In the first iteration, the decay parameters are initialized with random values from the search space. The algorithm proceeds by repeatedly sampling each decay parameter in turn.

In our evaluation, we use this framework to fit models of orders $K=1, 2, 3$ to the measured EDFs. As suggested in previous work \cite{XiangEtAl2011BayesianCharacterizationOfMultiSlopeDecays}, we apply the \replaced{Bayesian information criterion~(BIC)}{inverse Bayesian information criterion~(IBIC)} to determine the lowest possible model order that fits the data well. \added{Although it differs from the regular BIC by a negative sign, the definitions are sometimes used interchangeably.} By choosing the model with the maximum \replaced{BIC}{IBIC}, the Bayesian framework balances the degree of fit and a potential over-parameterization. The \replaced{BIC}{IBIC} for the model of order $K$ is given as \cite{XiangEtAl2011BayesianCharacterizationOfMultiSlopeDecays}
\begin{linenomath*}
	\begin{equation}
		\textrm{IBIC}_K = 2\, \textrm{ln}(\hat{\ell}_K) - (2K + 1) \, \textrm{ln} (L) \,,
	\end{equation}
\end{linenomath*}
where $\hat{\ell}_K$ is the maximum-likelihood determined with the slice-sampling algorithm, and $L$ is the number of EDF samples.

\subsection{Datasets}
We use two large RIR datasets for our evaluation. The first dataset is the \emph{Motus dataset} \cite{Goetz2021MotusDatasetPaper,Goetz2021MotusDatasetZenodo}, which contains 3320 RIRs measured in various acoustic conditions. More precisely, the RIRs were measured in a single room (approx. volume of \SI{60}{\metre^3}), where the furniture amount and placement were varied between measurements to generate 830 unique furniture combinations. The dataset features reverberation times between \SI{0.5}{\second} and \SI{2}{\second} at \SI{1}{\kilo\Hz}. Due to the complex geometries, the dataset also features acoustic wave phenomena like scattering. While most of its RIRs have a single-slope characteristic, only approximately $40$ RIRs have pronounced multi-slope EDFs due to non-uniform absorption material distributions.

The second dataset is the \emph{Room Transition dataset} \cite{McKenzie2021AcousticAnalysisDatasetRoomTransitionsPaper,McKenzie2021AcousticAnalysisDatasetRoomTransitionsDataset}. It contains measurements of four room transitions with a spatial resolution of \SI{5}{\centi\metre}, including positions with occluded line-of-sight between source and receiver. Due to the room coupling, a large number of EDFs in the dataset exhibit a multi-slope characteristic, where the amplitudes of the slopes vary considerably with the receiver position between the rooms. Furthermore, it was shown that the Room Transition dataset features complex acoustic phenomena of coupled room transitions, such as the portaling effect and distinctive direct-to-reverberant ratio transitions \cite{McKenzie2021AcousticAnalysisDatasetRoomTransitionsPaper}. Our evaluation excluded the transition ``Office to anechoic chamber'', thus using only 1212 RIRs of the dataset. The remaining room transitions are ``Meeting room to hallway'', ``Office to kitchen'', and ``Office to stairwell''. We cut away the last \SI{0.1}{\second} of all RIRs because they include a Hanning fade-out window that disturbs the fitting process.

Both datasets contain higher-order Ambisonic RIR. The following analyses are based on the omnidirectional channel and the six octave bands from \SIrange{125}{4000}{\Hz}. A preliminary analysis of the estimated \modelNest{} values showed that both datasets have similar average signal-to-noise ratios of approximately \SI{75}{\decibel}.
\subsection{Results}
\label{subsec:results}
We evaluated the DecayFitNet with respect to its fitting performance and computational complexity. The results are presented in the following section. 

\subsubsection{Example fits}
\begin{figure*}
	\centering
	\figline{\fig{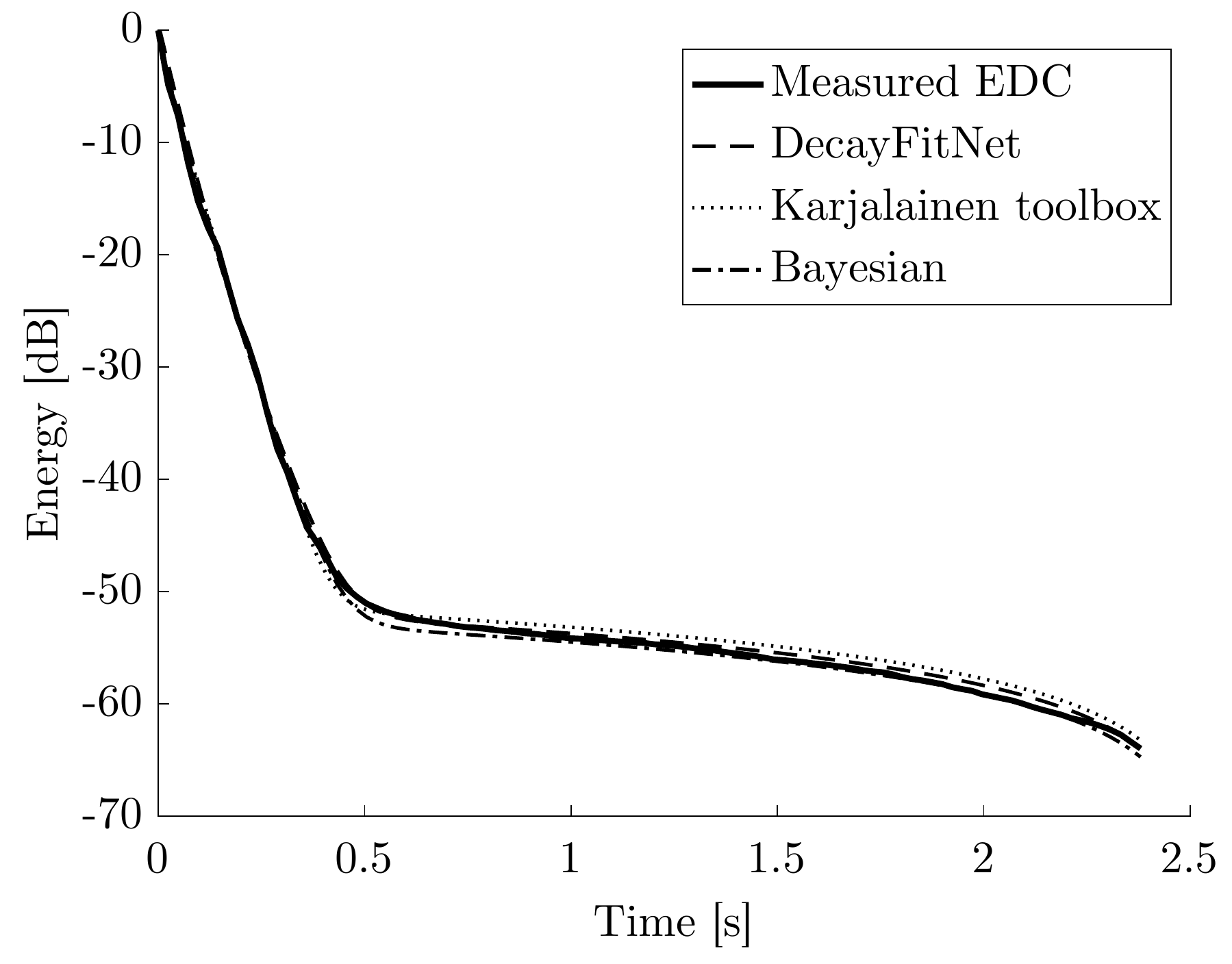}{0.45\textwidth}{(a)}\label{subfig:examplefitsingle}
		\fig{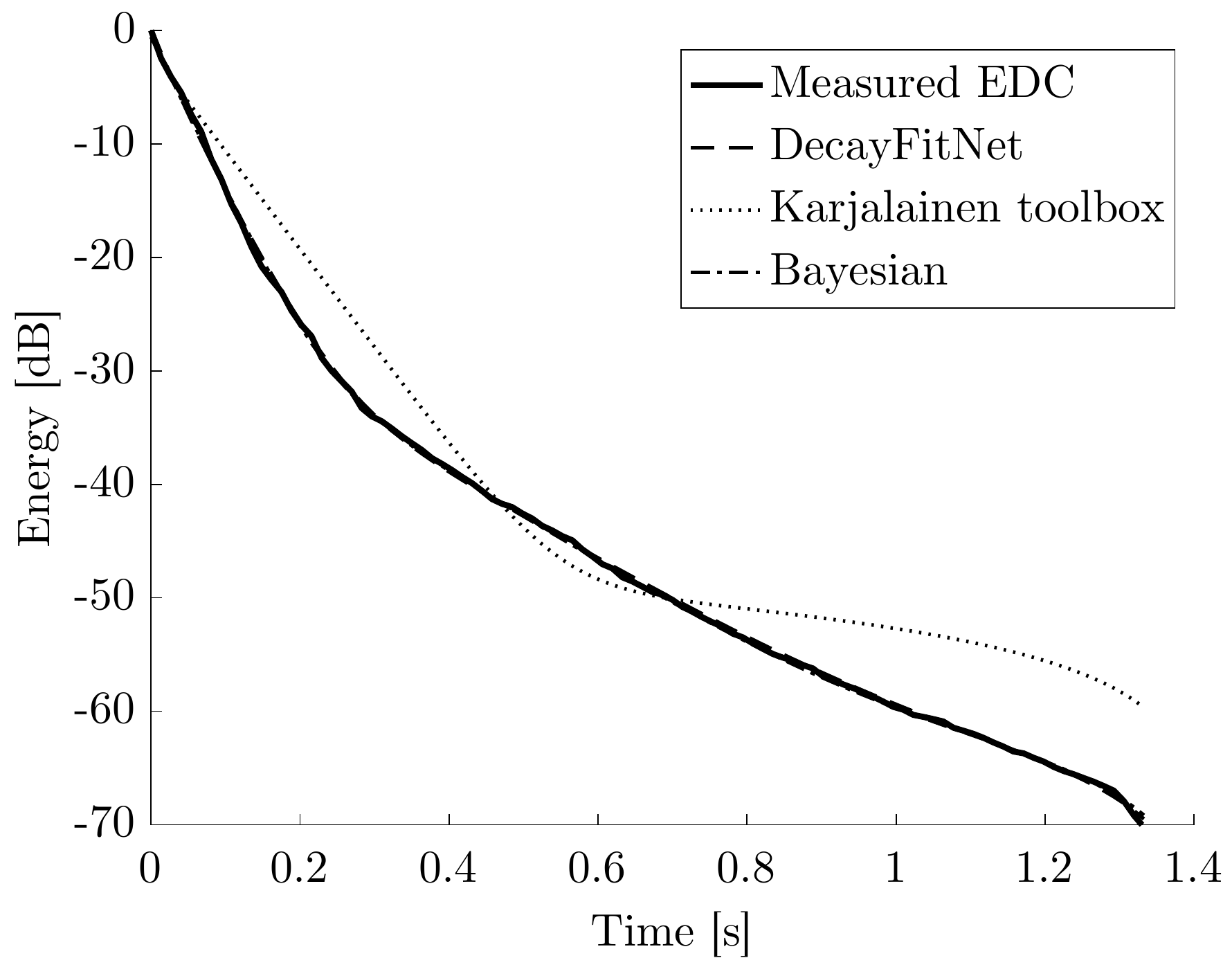}{0.45\textwidth}{(b)}\label{subfig:examplefitmulti}}
	\caption{Examples of fitting (a) a single-slope (from Motus dataset: measurement \#$500$, \mbox{loudspeaker $1$,} \SI{1}{\kilo\Hz} octave-band) and (b) a multi-slope EDF (from Room Transition dataset: Meeting room to hallway, source in room, no line of sight, \SI{25}{\centi\metre} position, \SI{1}{\kilo\Hz} octave-band). The measured EDFs are fitted with the proposed DecayFitNet, the grid-search variant of the Karjalainen toolbox \cite{Karjalainen2002EstimationOfModalDecayParameters}, and our implementation of the slice-sampling-based Bayesian decay analysis \cite{JasaXiang2009EfficientEstimationDecayParamsSliceSampling}. Analogously to the rest of our evaluation, the last \SI{5}{\percent} of EDF samples are excluded from the plot.}
	\label{fig:examplefit}
\end{figure*}

Figure \ref{fig:examplefit} shows example fits obtained with the proposed DecayFitNet for two measured EDFs. The figure also includes the corresponding fits obtained with the previously described grid search variant of the Karjalainen toolbox and our implementation of the slice-sampling-based Bayesian decay analysis. All approaches fit the single-slope EDF of Figure \ref{subfig:examplefitsingle} equally well. Figure \ref{subfig:examplefitmulti} depicts the resulting fits for a multi-slope EDF. The fits obtained with the proposed DecayFitNet architecture and the Bayesian decay analysis show good agreement with the measured EDF. In contrast, the Karjalainen toolbox is based on a single exponential plus noise model, thus being unable to fit EDFs with more than one slope. In the depicted scenario, the Karjalainen toolbox returns a slope between the two distinct slopes. Additionally, it overestimates the noise floor.

\subsubsection{Error analysis}
\begin{table*}[]
	\centering
	\caption{Median and \SI{99}{\percent} quantiles of the mean squared error between analyzed EDF and estimated fits. We compare the proposed DecayFitNet with the publicly available toolbox by Karjalainen et al. \cite{Karjalainen2002EstimationOfModalDecayParameters} and our own implementation of the slice-sampling-based Bayesian decay analysis \cite{JasaXiang2009EfficientEstimationDecayParamsSliceSampling}. The different fitting approaches are evaluated on two large, publicly available datasets.}
	\label{tbl:MSEcomparison}
	\begin{ruledtabular}
		\begin{tabular}{lcccccccc}
			\textbf{Dataset} & \multicolumn{2}{c}{\textbf{\begin{tabular}[c]{@{}c@{}}Karjalainen\\ (standard)\end{tabular}}} & \multicolumn{2}{c}{\textbf{\begin{tabular}[c]{@{}c@{}}Karjalainen\\ (grid)\end{tabular}}} & \multicolumn{2}{c}{\textbf{\begin{tabular}[c]{@{}c@{}}Bayesian\\ (50 iterations)\end{tabular}}} &  \multicolumn{2}{c}{\textbf{DecayFitNet}} \\ \cmidrule(lr){2-3} \cmidrule(lr){4-5} \cmidrule(lr){6-7} \cmidrule(lr){8-9}
			& median                   & \SI{99}{\percent} q.                   & median                  & \SI{99}{\percent} q.                  & median  & \SI{99}{\percent} q. & median  & \SI{99}{\percent} q. \\ \cmidrule(lr){1-9}
			Motus             & \SI{0.68}{\decibel}                 & \SI{6.69}{\decibel}                            & \SI{0.56}{\decibel}                    & \SI{5.32}{\decibel}           & \SI{0.23}{\decibel}  &     \SI{1.11}{\decibel}           & \SI{0.27}{\decibel}    & \SI{1.74}{\decibel}          \\
			Room Transition  & \SI{0.89}{\decibel}                     & \SI{45.56}{\decibel}                            & \SI{0.73}{\decibel}                    & \SI{22.53}{\decibel}          & \SI{0.35}{\decibel}  &    \SI{1.89}{\decibel}                & \SI{0.47}{\decibel}    & \SI{2.75}{\decibel}        \\ 
		\end{tabular}
	\end{ruledtabular}
\end{table*}

\explain{Slice-sampling-based decay analysis was added to caption of Table I.}Table \ref{tbl:MSEcomparison} summarizes the fitting results on the entire datasets. The table shows medians and \SI{99}{\percent} quantile values of the mean squared error (MSE) between measured octave-band filtered EDFs and obtained fits. The last \SI{5}{\percent} of EDF samples are excluded from the MSE calculation, following the reasoning described in Section \ref{subsec:preprocessing}. Furthermore, $11$ RIRs of the Motus dataset were excluded from the analysis because they featured transient noise artifacts. Such artifacts introduce large discontinuities into the corresponding EDFs, which violate the model of Eq. (\ref{eq:MultiSlopeModel}). Consequently, all tested fitting approaches had large MSEs for EDFs with such artefacts, allowing us to detect and exclude these measurements.\\
The table indicates that the Karjalainen toolbox, its grid-search variant, the slice-sampling-based Bayesian analysis, and the proposed DecayFitNet accurately fit the Motus dataset EDFs, with median MSEs of \SI{0.68}{\decibel} or lower. However, the \SI{99}{\percent} quantile values indicate that the spread of achieved MSE values is higher for the Karjalainen toolbox than for the proposed DecayFitNet and the Bayesian analysis. This suggests that the proposed DecayFitNet achieves slightly more robust fitting on a dataset with mostly single-slope EDFs than a well-established single-slope fitting toolbox. Furthermore, its performance \added{in terms of median errors} is comparable to an existing multi-slope fitting approach. \added{However, the DecayFitNet is slightly less robust than the Bayesian decay analysis as indicated by the increased \SI{99}{\percent} quantile value.}

The results for the Room Transition dataset show a similar trend. While median values do not differ considerably between the four approaches, the standard Karjalainen toolbox and its grid-search variant exhibit increased \SI{99}{\percent} quantile values of \SI{45.56}{\decibel} and \SI{22.53}{\decibel}, respectively. This considerable variability in fitting performance can be attributed to the insufficient model order for fitting multi-slope EDFs. In contrast, the \SI{99}{\percent} quantile value for fitting the Room Transition dataset with the proposed DecayFitNet is \SI{2.75}{\decibel} and therefore only slightly higher than for the Motus dataset. The Bayesian analysis achieves a similar performance\added{, although being a bit more robust as indicated by the lower \SI{99}{\percent} quantile value}. Nevertheless, this result suggests that the DecayFitNet can robustly fit large quantities of multi-slope EDFs. 

\begin{figure*}
	\centering
	\figline{\fig{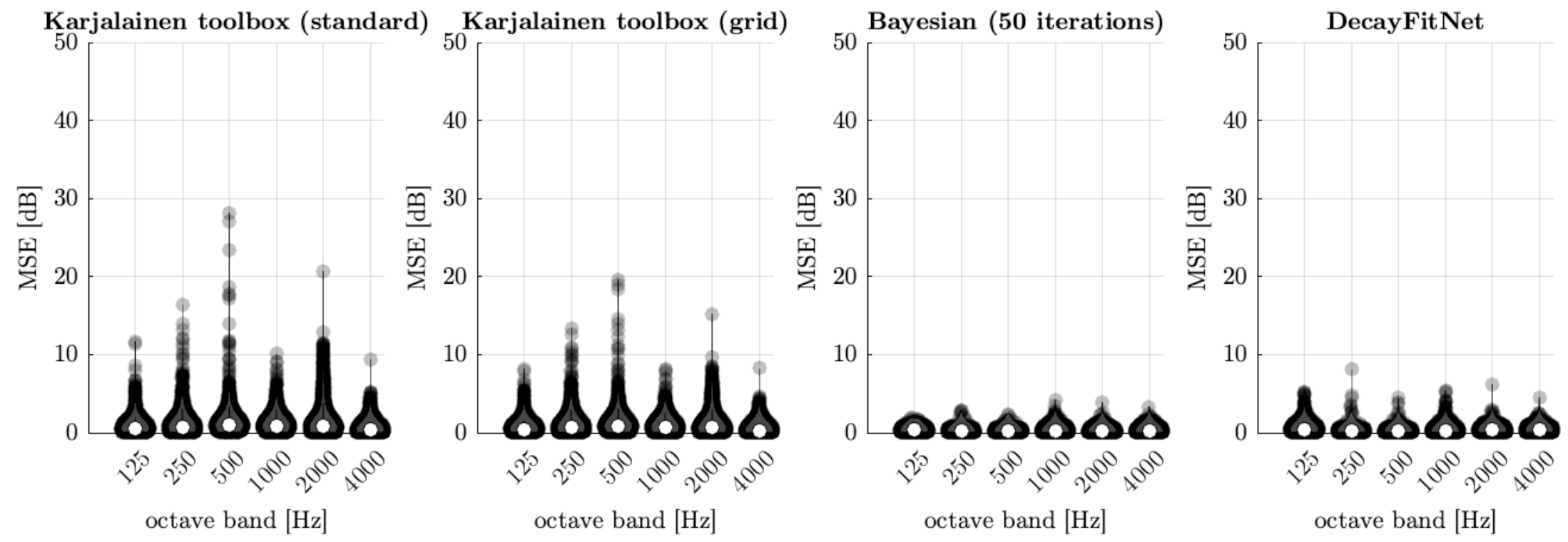}{\textwidth}{(a)}\label{subfig:violinmotus}}
	\figline{\fig{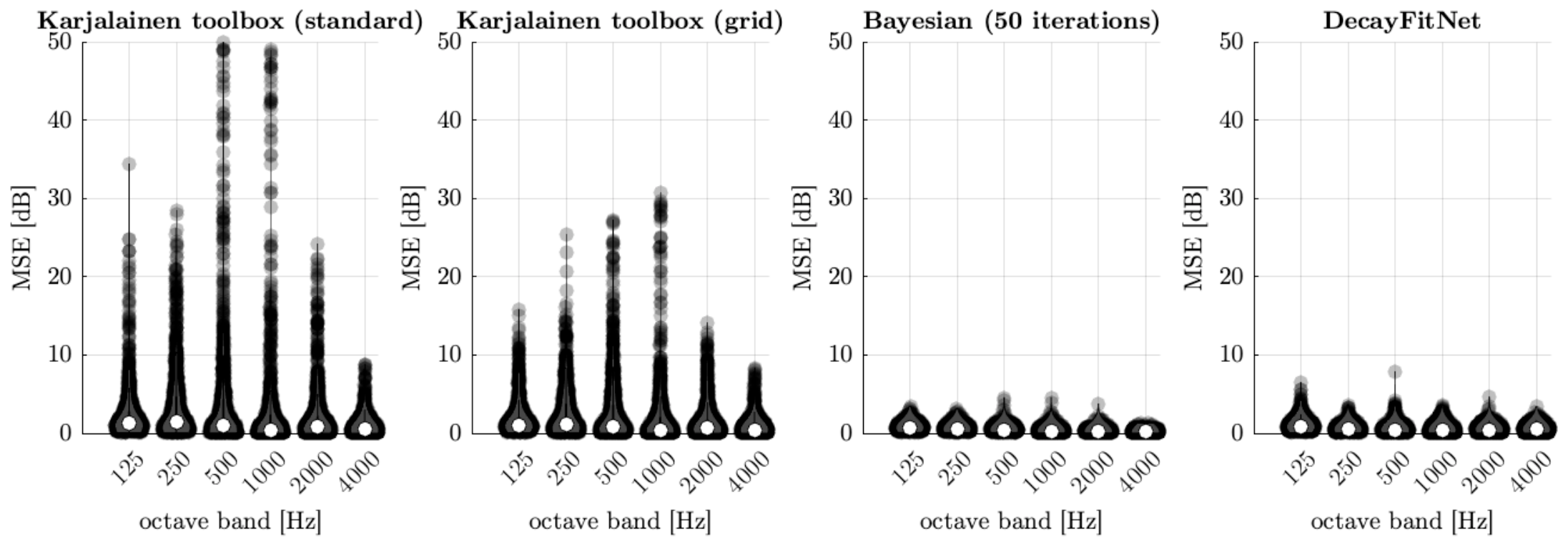}{\textwidth}{(b)}\label{subfig:violinroomtransition}}
	\caption{Violin plots of the mean squared error between measured octave-band-filtered EDFs and the corresponding fits obtained with different approaches. The evaluation is based on the (a) Motus dataset and the (b) Room Transition dataset. Light gray circles indicate individual data points, and white circles indicate the median. MSE values greater than \SI{50}{\decibel} have been excluded from the plot for clarity.}
	\label{fig:violin}
\end{figure*}

Figure \ref{fig:violin} supports these findings. It shows violin plots of the MSE values that were the basis for the calculations of \mbox{Table \ref{tbl:MSEcomparison}.} Figure \ref{subfig:violinmotus} shows that all approaches achieve low MSE values for the Motus dataset across all frequency bands. Most MSE values lie below \SI{10}{\decibel}, and the spread of values below \SI{10}{\decibel} is slightly bigger for the Karjalainen toolbox than for the DecayFitNet and the Bayesian analysis. Figure \ref{subfig:violinroomtransition} shows the results for the Room Transition dataset. The plots exhibit larger MSE values for the two Karjalainen toolbox approaches, with many data points well above \SI{10}{\decibel}. A more thorough analysis of the high MSE values revealed that they can be attributed to multi-slope EDFs, for which the model order of the Karjalainen toolbox is too low. In contrast, the proposed DecayFitNet and the Bayesian analysis achieve low fitting errors for the Room Transition dataset, with all MSE values below \SI{10}{\decibel}. \added{The Bayesian decay analysis exhibits slightly reduced MSE value spreads compared to the DecayFitNet.}

\subsubsection{Computational efficiency analysis}
The execution time of the DecayFitNet depends largely on the number of model parameters, i.e., the learned weights and biases of the different layers. With approximately \num{677000} parameters, the DecayFitNet is rather lightweight and could run on mobile devices.

Table \ref{tbl:Efficiency} compares the computational efficiency of the evaluated decay analysis approaches. It shows how much time the different approaches require to analyze the entire Motus dataset of \num{19920}~EDFs (\num{3320} RIRs $\times\ 6$ octave bands). The evaluation was carried out on a modern laptop computer (model year 2019) without a dedicated graphics processing unit (GPU). Whereas the standard version of the Karjalainen toolbox required almost two hours for the analysis, the Bayesian analysis and the proposed DecayFitNet were considerably faster. For the latter two approaches, we averaged the elapsed time of $100$ runs over the entire dataset to account for temporary drops of processing power that could bias the result. \replaced{The table shows that the DecayFitNet achieves an almost 30-fold speed-up compared to the Bayesian analysis. This speed-up can be attributed to the lightweight structure of the DecayFitNet and the iterative nature of the slice-sampling-based Bayesian analysis.}{The table shows that the neural-network-based analysis can process a large dataset with \num{19920} EDFs within seconds, which can be attributed to the lightweight structure of the DecayFitNet. However, it is important to note that the presented numbers should only be understood as an orientation for the reader. It is obvious that optimized implementations are an interesting engineering task for future work, which could reduce the computation times of all approaches even further. For example, due to the CNN-based architecture, our approach can easily be scaled up and parallelized to efficiently process large amounts of data with modern GPUs.}

\deleted{Apart from the execution times, Table \ref{tbl:Efficiency} also features two measures that provide additional insights into the computational efficiency of the approaches on other machines. The execution time of the slice-sampling-based Bayesian analysis depends largely on how fast the algorithm converges to the parameter space region surrounding the maximum likelihood. Therefore, we determined the average number of likelihood evaluations for the analysis of a single EDF. This measure indicates how many steps the algorithm takes through the parameter space until a satisfactory solution is found.}

\begin{table}[]
	\centering
	\caption{Runtime efficiency comparison of the evaluated decay analysis algorithms. For the Bayesian decay analysis and the DecayFitNet, the elapsed time for analyzing the entire Motus dataset (\num{19920} EDFs in total) is calculated as the average over $100$ runs.}
	\label{tbl:Efficiency}
	\begin{ruledtabular}
		\begin{tabular}{lll}
			\textbf{Karjalainen (standard)} &                                                  &                       \\
			& \SI{6281}{\second} (\SI{1.75}{\hour})                     \\ \hline
			\textbf{Bayesian}                                                      &                                                  &                       \\
			& \SI{733}{\second} (\SI{12.2}{\minute}) \\ \hline
			\textbf{DecayFitNet}                                                   &                                                  &                       \\
			& \SI{26.6}{\second}               \\       
		\end{tabular}
	\end{ruledtabular}
\end{table}

\section{Discussion}
\label{sec:Discussion}
The results presented in Section \ref{subsec:results} indicate that the proposed neural network structure can robustly fit single-slope and multi-slope EDFs on large datasets without prior parameter tuning or supervision. 

Our evaluation was based on two large datasets of more than $1000$ RIRs each, corresponding to more than \num{20000} EDFs across $6$ octave bands in total. They feature various acoustic conditions, such as varying amounts and placements of absorptive materials, diffraction and scattering from the room geometry, and room coupling. On both evaluated datasets, the DecayFitNet and the Bayesian decay analysis outperform the toolbox by \mbox{Karjalainen et al. \cite{Karjalainen2002EstimationOfModalDecayParameters},} which is a well-established toolbox for fitting EDFs with a single slope and a noise term. It is important to note that the Karjalainen toolbox cannot fit multi-exponential decays, thus explaining the degraded performance on the Room Transition dataset, which contains many multi-slope EDFs. In contrast, the Karjalainen toolbox performed well on the Motus dataset, which primarily contains single-slope decays. Nevertheless, it exhibited a \added{considerably} larger spread of errors than the DecayFitNet and the Bayesian decay analysis. The latter two approaches performed similarly well on both datasets\added{, although the Bayesian analysis is slightly more robust as indicated by the lower error spreads}. 


Furthermore, it is interesting to note that the DecayFitNet is completely deterministic at \replaced{runtime}{inference time}, because it always applies the same set of parameters to its inputs. This property limits the number of executed operations to a fixed value and ensures that the same results are obtained for repeated runs. In contrast, iterative algorithms like slice-sampling exhibit some degree of uncertainty, depending on the random steps taken throughout the execution. This uncertainty affects the consistency of results and the execution time alike. 

To the best of the authors' knowledge, at this point, there is no study examining the performance of state-of-the-art decay analysis algorithms on large amounts of data. We found that the slice-sampling-based Bayesian analysis performs well on large-scale datasets, despite the iterative nature of the approach. As a fully-deterministic alternative approach, we have proposed a neural network structure. With the presented DecayFitNet, potential numerical difficulties have moved from the application (and the user) to the network training stage. The network training has to be performed only once, and the pre-trained network can subsequently be used by users without any additional effort. This shift decreases the required user tuning and also reduces the computational complexity at \replaced{run-time}{inference time}. The efficiency and low user tuning of the neural network structure are balanced by the decreased possibilities of adjusting an incorrectly fitted EDF. While other approaches have adjustable parameters to fix a wrong EDF fit, the output of the neural network cannot be changed unless it is trained with new data. 

Lastly, we found that the presented DecayFitNet \replaced{achieves an almost 30-fold speed-up compared to the slice-sampling-based Bayesian decay analysis.}{can analyze a large dataset of almost \num{20000} EDFs in less than \SI{30}{\second}}. \replaced{This}{The high} computational efficiency \deleted{improvement} is an important step toward bringing decay analysis algorithms to mobile devices. Furthermore, it benefits the processing of large datasets, where GPU-accelerated machines can leverage the full potential of our CNN-based architecture. In principle, the Bayesian formulation allows the user to find a satisfactory fit for every EDF, as long as the parameter space is searched long enough. In some cases, this may require increasing the number of iterations in the slice-sampling algorithm, thus also increasing the computational costs. In our experiments, we found that $50$ iterations sufficiently explore the parameter space to yield good EDF fits in most cases\added{, whereas the fitting performance degraded quickly with a decreasing number of iterations}. \deleted{While the fitting performance degraded quickly with a decreasing number of iterations, the execution time remained high compared to the proposed neural network approach. This can be explained by the iterative exploration of the parameter space, where the first iterations evaluate the likelihood function more often than latter iterations.}

\section{Decay Analysis Toolbox}
\label{sec:Toolbox}

As part of this work, we provide an open source toolbox for both Python and Matlab. The toolbox includes a pretrained neural network model, a code interface to interact with the model, our implementation of the slice-sampling-based Bayesian decay analysis, as well as some utility functions. The main goal of the toolbox is to present a simple package to estimate multi-slope EDFs that can be used with minimal code. Although a pretrained model is provided, the toolbox also includes the ability to train and export a model using a custom dataset. This could be useful in scenarios where the generalization of the pretrained model is not accurate enough. The source code and documentation for the toolbox are available at \url{https://github.com/georg-goetz/DecayFitNet/}.

\begin{figure*}[ht]
	\begin{center}\includegraphics[width=0.85\textwidth]{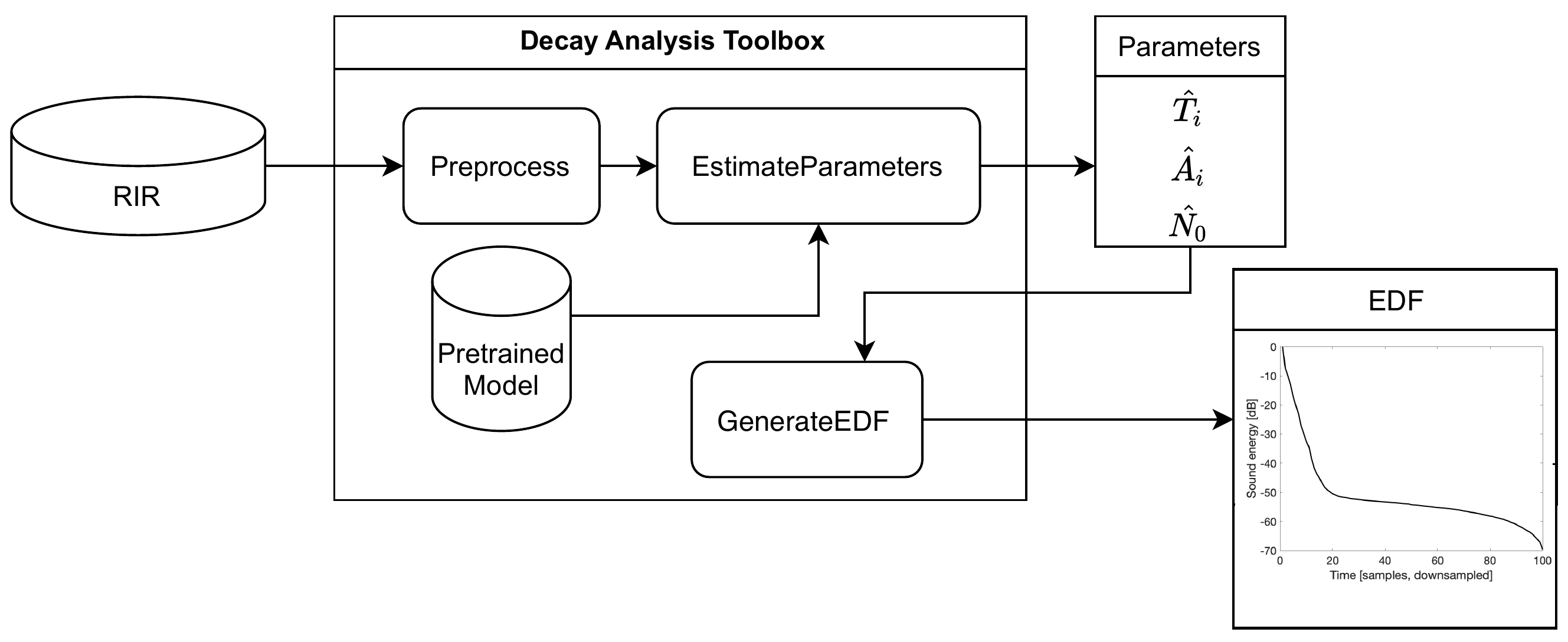}
		\caption{\label{fig:toolbox} Workflow and structure of the Decay Analysis Toolbox to estimate the EDF parameters from a RIR.}
	\end{center}
\end{figure*}

The structure and workflow of the toolbox are presented in Figure \ref{fig:toolbox}. The main functionality includes:
\begin{itemize}
	\item Impulse responses pre-processing for the neural-network-based decay analysis (as described in Section  \ref{subsec:preprocessing}) .
	\item Estimation of parameters \modelT, \modelA, and \modelN{} from Eq. (\ref{eq:MultiSlopeModel}) by doing a forward pass of the pre-trained model, or by applying the slice-sampling-based Bayesian decay analysis.
	\item Generation of the fitted EDF using the estimated parameters.
	\item Training and export of the DecayFitNet neural network model.
\end{itemize}

The toolbox exports the neural network model and operates using the Open Neural Network Exchange format (ONNX) \cite{ONNX}. ONNX is an open and common format that allows for fully-trained machine learning models to be distributed and utilised by a variety of frameworks and platforms. While the toolbox only includes interfaces for Python and Matlab, a trained model exported to the ONNX format can be supported by many applications, requiring only the porting of the pre-processing and EDF generation code. The provided model was trained using the synthetic dataset described in Section \ref{subsec:dataset}.

\section{Conclusions}
\label{sec:conclusions}
This paper proposed a neural network structure for fitting multi-exponential sound energy decay functions (EDFs). It was shown that such a network can be trained on a dataset of synthetically generated EDFs. We presented the DecayFitNet as an example architecture of the proposed approach. A large-scale evaluation applied the DecayFitNet and two comparable state-of-the-art methods on two large datasets of real-world measurements with more than $1000$ RIRs each, corresponding to over \num{20000} EDFs across $6$ octave bands. The analyzed datasets featured various acoustic conditions, such as varying amounts of absorptive material, diffraction and scattering from the room geometry, and room coupling. 

The results of our evaluation indicate that the proposed neural network structure can robustly fit single-slope and multi-slope EDFs without prior parameter tuning or supervision by the user. Additionally, the presented DecayFitNet is \added{fully-deterministic during inference time} and computationally efficient, being capable of analyzing almost \num{20000} EDFs in less than $30$ seconds on a modern laptop computer~(2019). Our evaluation indicates that the DecayFitNet \replaced{achieves an almost 30-fold speed-up compared to state-of-the-art multi-slope decay analysis algorithms}{robustly estimates the model parameters from large datasets of measured EDFs}, \replaced{while exhibiting a comparably}{while achieving a comparable but slightly less} robust fitting performance \added{compared to state-of-the-art multi-slope decay analysis algorithms}. A decay analysis toolbox has been made publicly available for the audio community. 

The DecayFitNet and its corresponding toolbox may benefit future research on room acoustic analysis and modeling. Data-heavy machine-learning-based approaches may leverage its full potential regarding computational efficiency and robustness. 

\section*{Acknowledgments}
The project has received funding from from the Academy of Finland, project no. 317341. We acknowledge the computational resources provided by the Aalto Science-IT project.

\bibliography{decayfitnet}

\begin{thebibliography}{10}
\def\enquote#1,{``#1,''}
\def\enxquote#1{``#1''}
\expandafter\ifx\csname url\endcsname\relax
  \def\url#1{\texttt{#1}}\fi
\expandafter\ifx\csname urlprefix\endcsname\relax\def\urlprefix{URL }\fi
\providecommand{\bibinfo}[2]{#2}
\def\plainquote#1{``#1''}
\providecommand{\noopsort}[1]{}
\providecommand{\switchargs}[2]{#2#1}
\providecommand{\dourl}[1]{\href{http://#1}{\nolinkurl{#1}}}
  \def\eatspace #1{#1}

\bibitem{Kuttruff2000RoomAcoustics}
\bibinfo{author}{H.~Kuttruff}, \emph{\bibinfo{title}{{Room Acoustics}}},
  \bibinfo{}{4th} ed.  (\bibinfo{publisher}{Spon Press},
  \bibinfo{address}{London, UK}, \bibinfo{year}{2000}).

\bibitem{Schroeder1965BackwardsIntegration}
\bibinfo{author}{M.~R. Schroeder}, \enquote{\bibinfo{title}{{New Method of
  Measuring Reverberation Time}}},  \bibinfo{journal}{J. Acoust. Soc. Am.}
  \textbf{37}(3), \bibinfo{pages}{409--412} (\bibinfo{year}{1965})
  \dodoi{10.1121/1.1909343}.

\bibitem{ISO3382RoomAcousticMeasurementPart1Performance}
\bibinfo{author}{{ISO 3382-1}}, \enquote{\bibinfo{title}{{Acoustics -
  Measurement of room acoustic parameters - Part 1: Performance Spaces}}},
  \bibinfo{type}{Standard}, \bibinfo{institution}{International Organization
  for Standardization (ISO)}, \bibinfo{address}{Geneva, Switzerland}
  (\bibinfo{year}{2009}).

\bibitem{ISO3382RoomAcousticMeasurementPart2Ordinary}
\bibinfo{author}{{ISO 3382-2}}, \enquote{\bibinfo{title}{{Acoustics -
  Measurement of room acoustic parameters - Part 2: Reverberation time in
  ordinary rooms}}}, \bibinfo{type}{Standard},
  \bibinfo{institution}{International Organization for Standardization (ISO)},
  \bibinfo{address}{Geneva, Switzerland} (\bibinfo{year}{2008}).

\bibitem{Chu1978ComparisonOfReverberationMeasurementsUsingSchroederAndAveraging}
\bibinfo{author}{W.~T. Chu}, \enquote{\bibinfo{title}{{Comparison of
  reverberation measurements using Schroeder’s impulse method and
  decay‐curve averaging method}}},  \bibinfo{journal}{J. Acoust. Soc. Am.}
  \textbf{63}(5), \bibinfo{pages}{1444--1450} (\bibinfo{year}{1978})
  \dodoi{10.1121/1.381889}.

\bibitem{Morgan1997ParametricErrorAnalysisOfBackwardIntegrationLimit}
\bibinfo{author}{D.~R. Morgan}, \enquote{\bibinfo{title}{{A parametric error
  analysis of the backward integration method for reverberation time
  estimation}}},  \bibinfo{journal}{J. Acoust. Soc. Am.} \textbf{101}(5),
  \bibinfo{pages}{2686--2693} (\bibinfo{year}{1997}) \dodoi{10.1121/1.418557}.

\bibitem{Lundeby1995UncertaintiesMeasurementsInRoomAcoustics}
\bibinfo{author}{A.~Lundeby}, \bibinfo{author}{T.~E. Vigran},
  \bibinfo{author}{H.~Bietz}, and \bibinfo{author}{M.~Vorländer},
  \enquote{\bibinfo{title}{{Uncertainties of Measurements in Room Acoustics}}},
   \bibinfo{journal}{Acta Acustica united with Acustica} \textbf{81}(4),
  \bibinfo{pages}{344--355} (\bibinfo{year}{1995}).

\bibitem{GuskiVorlaender2013MeasurementUncertaintiesRTcausedbyNoise}
\bibinfo{author}{M.~Guski} and \bibinfo{author}{M.~Vorländer},
  \enquote{\bibinfo{title}{{Measurement Uncertainties of Reverberation Time
  caused by Noise}}}, in \emph{\bibinfo{booktitle}{Proceedings of the
  International Conference on Acoustics (ICA), the 39th annual congress of
  DEGA, and the 40th annual congress of AIA}}, \bibinfo{publisher}{DEGA},
  \bibinfo{address}{Merano, Italy} (\bibinfo{year}{2013}), pp.
  \bibinfo{pages}{2067--2070}.

\bibitem{Xiang1995EvaluationOfRTNonlinearRegression}
\bibinfo{author}{N.~Xiang}, \enquote{\bibinfo{title}{{Evaluation of
  reverberation times using a nonlinear regression approach}}},
  \bibinfo{journal}{J. Acoust. Soc. Am.} \textbf{98}(4),
  \bibinfo{pages}{2112--2121} (\bibinfo{year}{1995}) \dodoi{10.1121/1.414460}.

\bibitem{Karjalainen2002EstimationOfModalDecayParameters}
\bibinfo{author}{M.~Karjalainen}, \bibinfo{author}{P.~Antsalo},
  \bibinfo{author}{A.~Mäkivirta}, \bibinfo{author}{T.~Peltonen}, and
  \bibinfo{author}{V.~Välimäki}, \enquote{\bibinfo{title}{{Estimation of
  Modal Decay Parameters from Noisy Response Measurements}}},
  \bibinfo{journal}{J. Audio Eng. Soc.} \textbf{50}(11),
  \bibinfo{pages}{867--878} (\bibinfo{year}{2002}).

\bibitem{Eyring1931ReverberationTimeMeasurementsCoupledRooms}
\bibinfo{author}{C.~F. Eyring}, \enquote{\bibinfo{title}{{Reverberation time
  measurements in coupled rooms}}},  \bibinfo{journal}{J. Acoust. Soc. Am.}
  \textbf{3}(2), \bibinfo{pages}{181--206} (\bibinfo{year}{1931})
  \dodoi{10.1121/1.1915555}.

\bibitem{Hunt1939AnalysisOfSounDecayInRectRooms}
\bibinfo{author}{F.~V. Hunt}, \bibinfo{author}{L.~L. Beranek}, and
  \bibinfo{author}{D.~Y. Maa}, \enquote{\bibinfo{title}{{Analysis of Sound
  Decay in Rectangular Rooms}}},  \bibinfo{journal}{J. Acoust. Soc. Am.}
  \textbf{11}(1), \bibinfo{pages}{80--94} (\bibinfo{year}{1939})
  \dodoi{10.1121/1.1916010}.

\bibitem{Nilsson2004DecayInRoomsNonDiffuseSoundfieldCeilingTreatment}
\bibinfo{author}{E.~Nilsson}, \enquote{\bibinfo{title}{{Decay Processes in
  Rooms with Non-Diffuse Sound Fields Part I: Ceiling Treatment with Absorbing
  Material}}},  \bibinfo{journal}{Building Acoustics} \textbf{11}(1),
  \bibinfo{pages}{39--60} (\bibinfo{year}{2004})
  \dodoi{10.1260/1351010041217220}.

\bibitem{Xiang2001EvaluationOfDecayTimesInCoupledSpacesBayesian}
\bibinfo{author}{N.~Xiang} and \bibinfo{author}{P.~M. Goggans},
  \enquote{\bibinfo{title}{{Evaluation of decay times in coupled spaces:
  Bayesian parameter estimation}}},  \bibinfo{journal}{J. Acoust. Soc. Am.}
  \textbf{110}(3), \bibinfo{pages}{1415--1424} (\bibinfo{year}{2001})
  \dodoi{10.1121/1.1390334}.

\bibitem{Xiang2005DecayInCoupledSpacesReliabilityAnalysisBayesianDecay}
\bibinfo{author}{N.~Xiang}, \bibinfo{author}{P.~M. Goggans},
  \bibinfo{author}{T.~Jasa}, and \bibinfo{author}{M.~Kleiner},
  \enquote{\bibinfo{title}{{Evaluation of decay times in coupled spaces:
  Reliability analysis of Bayeisan decay time estimation}}},
  \bibinfo{journal}{J. Acoust. Soc. Am.} \textbf{117}(6),
  \bibinfo{pages}{3707--3715} (\bibinfo{year}{2005}) \dodoi{10.1121/1.1903845}.

\bibitem{BalintMuralterEtAl2019BayesianDecayTimeEstimationAbsorptionMeasurements}
\bibinfo{author}{J.~Balint}, \bibinfo{author}{F.~Muralter},
  \bibinfo{author}{M.~Nolan}, and \bibinfo{author}{C.-H. Jeong},
  \enquote{\bibinfo{title}{{Bayesian decay time estimation in a reverberation
  chamber for absorption measurements}}},  \bibinfo{journal}{J. Acoust. Soc.
  Am.} \textbf{146}(3), \bibinfo{pages}{1641--1649} (\bibinfo{year}{2019})
  \dodoi{10.1121/1.5125132}.

\bibitem{Pu2011SoundDecayPatternsEnergyFeedbackCoupled}
\bibinfo{author}{H.~Pu}, \bibinfo{author}{X.~Qiu}, and
  \bibinfo{author}{J.~Wang}, \enquote{\bibinfo{title}{{Different sound decay
  patterns and energy feedback in coupled volumes}}},  \bibinfo{journal}{J.
  Acoust. Soc. Am.} \textbf{129}(4), \bibinfo{pages}{1972--1980}
  (\bibinfo{year}{2011}) \dodoi{10.1121/1.3553223}.

\bibitem{SuGul2019DiffusionEquationModelingSoundEnergyFlowCoupled}
\bibinfo{author}{Z.~Sü~Gül}, \bibinfo{author}{E.~Odabaş},
  \bibinfo{author}{N.~Xiang}, and \bibinfo{author}{M.~Çalışkan},
  \enquote{\bibinfo{title}{{Diffusion equation modeling for sound energy flow
  analysis in multi domain structures}}},  \bibinfo{journal}{J. Acoust. Soc.
  Am.} \textbf{145}(4), \bibinfo{pages}{2703--2717} (\bibinfo{year}{2019})
  \dodoi{10.1121/1.5095877}.

\bibitem{XiangEtAl2011BayesianCharacterizationOfMultiSlopeDecays}
\bibinfo{author}{N.~Xiang}, \bibinfo{author}{P.~Goggans},
  \bibinfo{author}{T.~Jasa}, and \bibinfo{author}{P.~Robinson},
  \enquote{\bibinfo{title}{{Bayesian characterization of multiple-slope sound
  energy decays in coupled-volume systems}}},  \bibinfo{journal}{J. Acoust.
  Soc. Am.} \textbf{129}(2), \bibinfo{pages}{741--752} (\bibinfo{year}{2011})
  \dodoi{10.1121/1.3518773}.

\bibitem{JasaXiang2009EfficientEstimationDecayParamsSliceSampling}
\bibinfo{author}{T.~Jasa} and \bibinfo{author}{N.~Xiang},
  \enquote{\bibinfo{title}{{Efficient estimation of decay parameters in
  acoustically coupled-spaces using slice sampling}}},  \bibinfo{journal}{J.
  Acoust. Soc. Am.} \textbf{126}(3), \bibinfo{pages}{1269--1279}
  (\bibinfo{year}{2009}) \dodoi{10.1121/1.3158934}.

\bibitem{Cabrera2016CalculatingReverberationTimeFromIRSoftwareComparison}
\bibinfo{author}{D.~Cabrera}, \bibinfo{author}{J.~Xun}, and
  \bibinfo{author}{M.~Guski}, \enquote{\bibinfo{title}{{Calculating
  Reverberation Time from Impulse Responses: A Comparison of Software
  Implementations}}},  \bibinfo{journal}{Acoustics Australia} \textbf{44}(2),
  \bibinfo{pages}{369--378} (\bibinfo{year}{2016})
  \dodoi{10.1007/s40857-016-0055-6}.

\bibitem{AlvarezMorales2016AcousticCharacterisationDifferentRoomAcousticSoftware}
\bibinfo{author}{L.~{Álvarez-Morales}}, \bibinfo{author}{M.~Galindo},
  \bibinfo{author}{S.~Girón}, \bibinfo{author}{T.~Zamarreño}, and
  \bibinfo{author}{R.~M. Cibrián}, \enquote{\bibinfo{title}{{Acoustic
  Characterisation by Using Different Room Acoustics Software Tools: A
  Comparative Study}}},  \bibinfo{journal}{Acta Acustica united with Acustica}
  \textbf{102}(3), \bibinfo{pages}{578--591} (\bibinfo{year}{2016})
  \dodoi{10.3813/aaa.918975}.

\bibitem{Katz2004RoundRobinRoomAcousticIRAnalysisSoftware}
\bibinfo{author}{B.~F.~G. Katz}, \enquote{\bibinfo{title}{{International Round
  Robin on Room Acoustical Impulse Response Analysis Software 2004}}},
  \bibinfo{journal}{Acoustics Research Letters Online} \textbf{5}(4),
  \bibinfo{pages}{158--164} (\bibinfo{year}{2004}) \dodoi{10.1121/1.1758239}.

\bibitem{ALAM2020302}
\bibinfo{author}{M.~Alam}, \bibinfo{author}{M.~Samad},
  \bibinfo{author}{L.~Vidyaratne}, \bibinfo{author}{A.~Glandon}, and
  \bibinfo{author}{K.~Iftekharuddin}, \enquote{\bibinfo{title}{Survey on deep
  neural networks in speech and vision systems}},
  \bibinfo{journal}{Neurocomputing} \textbf{417}, \bibinfo{pages}{302--321}
  (\bibinfo{year}{2020}) \dodoi{10.1016/j.neucom.2020.07.053}.

\bibitem{young2017recent}
\bibinfo{author}{T.~Young}, \bibinfo{author}{D.~Hazarika},
  \bibinfo{author}{S.~Poria}, and \bibinfo{author}{E.~Cambria},
  \enxquote{\bibinfo{title}{Recent trends in deep learning based natural
  language processing}}  (\bibinfo{year}{2017}) \bibinfo{note}{{Available
  online:} \url{https://arxiv.org/abs/1708.02709}}.

\bibitem{Purwins2019}
\bibinfo{author}{H.~Purwins}, \bibinfo{author}{B.~Li},
  \bibinfo{author}{T.~Virtanen}, \bibinfo{author}{J.~Schluter},
  \bibinfo{author}{S.-Y. Chang}, and \bibinfo{author}{T.~Sainath},
  \enquote{\bibinfo{title}{Deep learning for audio signal processing}},
  \bibinfo{journal}{{IEEE} Journal of Selected Topics in Signal Processing}
  \textbf{13}(2), \bibinfo{pages}{206--219} (\bibinfo{year}{2019})
  \dodoi{10.1109/jstsp.2019.2908700}.

\bibitem{grumiaux_2021:SurveySoundSource}
\bibinfo{author}{P.-A. Grumiaux}, \bibinfo{author}{S.~Kiti{\'c}},
  \bibinfo{author}{L.~Girin}, and \bibinfo{author}{A.~Gu{\'e}rin},
  \enxquote{\bibinfo{title}{A {{Survey}} of {{Sound Source Localization}} with
  {{Deep Learning Methods}}}}  (\bibinfo{year}{2021}) \bibinfo{note}{{Available
  online:} \url{https://arxiv.org/abs/2109.03465}}.

\bibitem{ACEchallenge}
\bibinfo{author}{J.~Eaton}, \bibinfo{author}{N.~D. Gaubitch},
  \bibinfo{author}{A.~H. Moore}, and \bibinfo{author}{P.~A. Naylor},
  \enquote{\bibinfo{title}{Estimation of room acoustic parameters: The {ACE}
  challenge}},  \bibinfo{journal}{IEEE/ACM Trans. Audio, Speech, Language
  Process.} \textbf{24}(10), \bibinfo{pages}{1681--1693} (\bibinfo{year}{2016})
  \dodoi{10.1109/TASLP.2016.2577502}.

\bibitem{fan2020fast}
\bibinfo{author}{Z.~Fan}, \bibinfo{author}{V.~Vineet},
  \bibinfo{author}{H.~Gamper}, and \bibinfo{author}{N.~Raghuvanshi},
  \enquote{\bibinfo{title}{{Fast Acoustic Scattering Using Convolutional Neural
  Networks}}}, in \emph{\bibinfo{booktitle}{Int. Conf. Acoust., Speech, Sig.
  Proc. (ICASSP)}}, \bibinfo{address}{Online conference}
  (\bibinfo{year}{2020}), pp. \bibinfo{pages}{171--175},
  \dodoi{10.1109/icassp40776.2020.9054091}.

\bibitem{Wirler2021Scattering}
\bibinfo{author}{S.~Wirler}, \bibinfo{author}{S.~J. Schlecht}, and
  \bibinfo{author}{V.~Pulkki}, \enquote{\bibinfo{title}{{Machine Learning Based
  Auralization of Rigid Sphere Scattering}}}, in
  \emph{\bibinfo{booktitle}{International Conference on Immersive and 3D Audio
  (I3DA)}}, \bibinfo{address}{Online conference} (\bibinfo{year}{2021}),
  \dodoi{10.1109/i3da48870.2021.9610951}.

\bibitem{9414743}
\bibinfo{author}{Z.~Fan}, \bibinfo{author}{V.~Vineet}, \bibinfo{author}{C.~Lu},
  \bibinfo{author}{T.~W. Wu}, and \bibinfo{author}{K.~McMullen},
  \enquote{\bibinfo{title}{Prediction of object geometry from acoustic
  scattering using convolutional neural networks}}, in
  \emph{\bibinfo{booktitle}{Int. Conf. Acoust., Speech, Sig. Proc. (ICASSP)}},
  \bibinfo{address}{Online conference} (\bibinfo{year}{2021}), pp.
  \bibinfo{pages}{471--475}, \dodoi{10.1109/ICASSP39728.2021.9414743}.

\bibitem{bianco_2019:MachineLearningAcoustics}
\bibinfo{author}{M.~J. Bianco}, \bibinfo{author}{P.~Gerstoft},
  \bibinfo{author}{J.~Traer}, \bibinfo{author}{E.~Ozanich},
  \bibinfo{author}{M.~A. Roch}, \bibinfo{author}{S.~Gannot}, and
  \bibinfo{author}{C.-A. Deledalle}, \enquote{\bibinfo{title}{Machine learning
  in acoustics: Theory and applications}},  \bibinfo{journal}{The Journal of
  the Acoustical Society of America} \textbf{146}(5),
  \bibinfo{pages}{3590--3628} (\bibinfo{year}{2019}) \dodoi{10.1121/1.5133944}.

\bibitem{GoodBengCour16}
\bibinfo{author}{I.~J. Goodfellow}, \bibinfo{author}{Y.~Bengio}, and
  \bibinfo{author}{A.~Courville}, \emph{\bibinfo{title}{Deep Learning}}
  (\bibinfo{publisher}{MIT Press}, \bibinfo{address}{Cambridge, MA, USA},
  \bibinfo{year}{2016}) \bibinfo{note}{{Available online:}
  \url{http://www.deeplearningbook.org}}.

\bibitem{FERNANDEZDELGADO201911}
\bibinfo{author}{M.~Fernández-Delgado}, \bibinfo{author}{M.~Sirsat},
  \bibinfo{author}{E.~Cernadas}, \bibinfo{author}{S.~Alawadi},
  \bibinfo{author}{S.~Barro}, and \bibinfo{author}{M.~Febrero-Bande},
  \enquote{\bibinfo{title}{An extensive experimental survey of regression
  methods}},  \bibinfo{journal}{Neural Networks} \textbf{111},
  \bibinfo{pages}{11--34} (\bibinfo{year}{2019})
  \dodoi{10.1016/j.neunet.2018.12.010}.

\bibitem{Scherer2010MaxPool}
\bibinfo{author}{D.~Scherer}, \bibinfo{author}{A.~Müller}, and
  \bibinfo{author}{S.~Behnke}, \enquote{\bibinfo{title}{{Evaluation of Pooling
  Operations in Convolutional Architectures for Object Recognition}}}, in
  \emph{\bibinfo{booktitle}{Proceedings of the International Conference on
  Artificial Neural Networks (ICANN), {Part III}}}, Lecture Notes in Computer
  Science, \bibinfo{publisher}{Springer}, \bibinfo{address}{Thessaloniki,
  Greece} (\bibinfo{year}{2010}), pp. \bibinfo{pages}{92--101},
  \dodoi{10.1007/978-3-642-15825-4\_10}.

\bibitem{Nair2010ReLUImproveRestrictedBoltzmannMachines}
\bibinfo{author}{V.~Nair} and \bibinfo{author}{G.~E. Hinton},
  \enquote{\bibinfo{title}{{Rectified Linear Units Improve Restricted Boltzmann
  Machines}}}, in \emph{\bibinfo{booktitle}{Proceedings of the 27th
  International Conference on Machine Learning (ICML-10)}},
  \bibinfo{address}{Haifa, Israel} (\bibinfo{year}{2010}), pp.
  \bibinfo{pages}{807--814}.

\bibitem{Pinkus1999ApproximationTO}
\bibinfo{author}{A.~Pinkus}, \enquote{\bibinfo{title}{Approximation theory of
  the {MLP} model in neural networks}},  \bibinfo{journal}{Acta Numerica}
  \textbf{8}, \bibinfo{pages}{143--195} (\bibinfo{year}{1999}).

\bibitem{lecun95convolutional}
\bibinfo{author}{Y.~Lecun} and \bibinfo{author}{Y.~Bengio},
  \enquote{\bibinfo{title}{Convolutional networks for images, speech and time
  series}}, in \emph{\bibinfo{booktitle}{The Handbook of Brain Theory and
  Neural Networks}},  edited by \bibinfo{editor}{M.~A. Arbib}
  (\bibinfo{publisher}{The MIT Press}, \bibinfo{year}{1995}), pp.
  \bibinfo{pages}{255--258}.

\bibitem{bilen2016}
\bibinfo{author}{H.~Bilen} and \bibinfo{author}{A.~Vedaldi},
  \enquote{\bibinfo{title}{Integrated perception with recurrent multi-task
  neural networks}}, in \emph{\bibinfo{booktitle}{Proceedings of the 30th
  International Conference on Neural Information Processing Systems}}, NIPS'16,
  \bibinfo{publisher}{Curran Associates Inc.}, \bibinfo{address}{Red Hook, NY,
  USA} (\bibinfo{year}{2016}), p. \bibinfo{pages}{235–243}.

\bibitem{KingmaBa2015Adam}
\bibinfo{author}{D.~P. Kingma} and \bibinfo{author}{J.~L. Ba},
  \enquote{\bibinfo{title}{{Adam: A Method for Stochastic Optimization}}}, in
  \emph{\bibinfo{booktitle}{International Conference on Learning
  Representations (ICLR)}}, \bibinfo{address}{San Diego, CA, USA}
  (\bibinfo{year}{2015}).

\bibitem{Loshchilov2019DecoupledWeightDecay}
\bibinfo{author}{I.~Loshchilov} and \bibinfo{author}{F.~Hutter},
  \enquote{\bibinfo{title}{{Decoupled Weight Decay Regularization}}}, in
  \emph{\bibinfo{booktitle}{International Conference on Learning
  Representations (ICLR)}}, \bibinfo{address}{New Orleans, LA, USA}
  (\bibinfo{year}{2019}).

\bibitem{Loshchilov2017CosAnnealingWarmRestarts}
\bibinfo{author}{I.~Loshchilov} and \bibinfo{author}{F.~Hutter},
  \enquote{\bibinfo{title}{{SGDR: Stochastic Gradient Descent with Warm
  Restarts}}}, in \emph{\bibinfo{booktitle}{International Conference on
  Learning Representations (ICLR)}}, \bibinfo{address}{Toulon, France}
  (\bibinfo{year}{2017}).

\bibitem{Goetz2021MotusDatasetPaper}
\bibinfo{author}{G.~Götz}, \bibinfo{author}{S.~J. Schlecht}, and
  \bibinfo{author}{V.~Pulkki}, \enquote{\bibinfo{title}{{A dataset of
  higher-order Ambisonic room impulse responses and 3D models measured in a
  room with varying furniture}}}, in \emph{\bibinfo{booktitle}{International
  Conference on Immersive and 3D Audio (I3DA)}}, \bibinfo{address}{Online
  conference} (\bibinfo{year}{2021}), pp. \bibinfo{pages}{1--8},
  \dodoi{10.1109/i3da48870.2021.9610933}.

\bibitem{Goetz2021MotusDatasetZenodo}
\bibinfo{author}{G.~Götz}, \bibinfo{author}{S.~J. Schlecht}, and
  \bibinfo{author}{V.~Pulkki}, \enxquote{\bibinfo{title}{{Motus: A dataset of
  higher-order Ambisonic room impulse responses and 3D models measured in a
  room with varying furniture (Version 1.0)}}}  (\bibinfo{year}{2021}),
  \dodoi{10.5281/zenodo.4923187} \bibinfo{note}{{Zenodo}}.

\bibitem{McKenzie2021AcousticAnalysisDatasetRoomTransitionsPaper}
\bibinfo{author}{T.~McKenzie}, \bibinfo{author}{S.~J. Schlecht}, and
  \bibinfo{author}{V.~Pulkki}, \enquote{\bibinfo{title}{{Acoustic Analysis and
  Dataset of Transitions Between Coupled Rooms}}}, in
  \emph{\bibinfo{booktitle}{Int. Conf. Acoust., Speech, Sig. Proc. (ICASSP)}},
  \bibinfo{address}{Online conference} (\bibinfo{year}{2021}), pp.
  \bibinfo{pages}{481--485}, \dodoi{10.1109/icassp39728.2021.9415122}.

\bibitem{McKenzie2021AcousticAnalysisDatasetRoomTransitionsDataset}
\bibinfo{author}{T.~McKenzie}, \bibinfo{author}{S.~J. Schlecht}, and
  \bibinfo{author}{V.~Pulkki}, \enxquote{\bibinfo{title}{{A dataset of measured
  spatial room impulse responses for the transition between coupled rooms
  (Version 1.2)}}}  (\bibinfo{year}{2021}), \dodoi{10.5281/zenodo.4636068}
  \bibinfo{note}{{Zenodo}}.

\bibitem{ONNX}
\bibinfo{author}{J.~Bai}, \bibinfo{author}{F.~Lu}, \bibinfo{author}{K.~Zhang},
  \emph{et~al.}, \enxquote{\bibinfo{title}{{ONNX}: Open neural network
  exchange}} , \bibinfo{howpublished}{\url{https://github.com/onnx/onnx}}
  (\bibinfo{year}{2019}).

\end{thebibliography}

\end{document}